\documentclass[twocolumn]{aastex62}
\usepackage{amsmath}

\newcommand{\fracbrac}[2]{\left(\frac{#1}{#2}\right)}
\newcommand{\scK}{{\mathcal K}}

\newcommand{\paren}[1]{\left( #1\right)}

\newcommand{\online}{\href{https://github.com/shadden/IntegrableModelforResonances}{online}}
\newcommand{\website}{\href{https://github.com/shadden/IntegrableModelforResonances}{github.com/shadden/IntegrableModelforResonances}}

\shorttitle{Resonances}
\shortauthors{Hadden}

\begin{document}
\title{An Integrable Model for the Dynamics of Planetary Mean Motion Resonances}
\author[0000-0002-1032-0783]{Sam~Hadden}
\affiliation{Harvard-Smithsonian Center for Astrophysics, 60 Garden St., MS 51, Cambridge, MA 02138, USA}
\correspondingauthor{Sam~Hadden}
\email{samuel.hadden@cfa.harvard.edu}
\begin{abstract}
I consider the dynamics of mean motion resonances between pairs of co-planar planets and derive a new integrable Hamiltonian model for planets' resonant motion.
The new model generalizes integrable Hamiltonians previously derived for first-order resonances to the case of higher-order resonances by exploiting a surprising near-symmetry of the full, non-integrable Hamiltonians of higher-order resonances.
Whereas past works have frequently relied on truncated disturbing function expansions to derive integrable approximations to resonant motion, I show that no such expansion is necessary to derive an integrable model. 
This enables the new model to accurately capture the dynamics of both first- and higher-order resonances for eccentricities up to orbit-crossing.
I demonstrate that predictions of the new integrable model agree well with numerical integrations of resonant planet pairs.
Finally, I explore the secular evolution of resonant planets' eccentricities. 
I show that the secular dynamics are governed by conservation of an AMD-like quantity.
I also demonstrate that secular frequencies depend on planets' resonant libration amplitude and this generally gives rise to a secular resonance inside the mean motion resonance at large libration amplitudes. 
The integrable model derived in this work can serve as a framework for analyzing the dynamics of planetary MMRs in a wide variety of contexts.
\end{abstract}

\keywords{celestial mechanics --- keyword2 --- keyword3}

\section{Introduction}
    Mean-motion resonances (MMRs) occur in a wide variety of contexts in celestial mechanics. Prominent examples in our own solar system include the 3:2 resonance between Neptune and Pluto, numerous resonances between giant planet satellites, and a wide variety of resonant phenomena in the asteroid and Kuiper belts \citep[e.g.,][]{Yoder1973,Peale1976,Peale1986,Peale1999,Greenberg1977,Malhotra1988,Malhotra1994}.
    A number of giant planets discovered by radial velocity (RV) surveys are found in or near MMRs \citep[e.g.,][]{Marcy2001,Godziewski2001,Lee2002,Johnson2011,Wittenmyer2014,Trifonov2014,Trifonov2019}.
    While MMRs are rare among multi-transiting super-Earths \citep{Fabrycky2014}, a handful of systems contain one or more resonant planet pair \citep[e.g.,][]{Gozdziewski2016,Mills2016,Migaszewski2017,Panichi2019} including TRAPPIST-1's 7-planet resonant chain \citep{Gillon2017,Luger2017}. 
    An analytic understanding of resonant dynamics is desirable given the prevalence of MMRs in solar system and exoplanetary contexts.

    Analytic treatments of resonant motion often rely on simplifying assumptions such as treating one of the planets as a massless test-particle, or assuming one of the orbits as (nearly) circular \citep[e.g.][]{Message1966,Henrard1983,Lemaitre1984,Peale1986,Malhotra1988,MD1999ssd}. These assumptions allow the resonant dynamics to be treated analytically because they yield equations of motion that depend on only one critical resonant angle. If, however, both planets have similar eccentricities then it is necessary to consider multiple critical resonant angles associated with various ``sub-resonances" that are potentially all of equal importance. This obstacle is circumvented when treating the dynamics of first-order MMRs to lowest order in eccentricities due to the existence of an additional constant of motion \citep{Sessin1984} which can be constructed by means of a canonical transformation \citep{wisdom1986,Henrard1986,FerrazMello2007,BatyginMorbidelli2013analytic}. 

    Beyond analytic treatments of MMRs in the circular restricted problem, a number of studies of MMRs have focused on computing families of periodic orbits (sometimes referred to as apsidal corotation resonances or ACRs) in the unrestricted planetary three-body problem that form the phase-space `skeleton' around which more general (quasi-)periodic resonant orbits are built \citep[e.g.,][]{Ferraz-Mello1993,Ferraz-Mello2003,Malhotra2002,Beauge2003,Lee2004,Hadjidemetriou2006,Michtchenko2006,Antoniadou2014}.
    Other studies have provided detailed explorations of the global dynamics of specific MMRs in the unrestricted problem, often focusing on a particular planetary or satellite system \citep[e.g.,][]{Michtchenko2001,Callegari2004,Callegari2006,Callegari2007,Michtchenko2008a,Michtchenko2008b}.
    \citet{Delisle2014} treat the dynamics of generic MMRs, both first- and higher order, and derive an integrable model for the motion in the vicinity of ACRs using the same canonical transformation that \citet{wisdom1986} and \cite{Henrard1986} employ for first-order resonances. The goal of this paper is to present a more general study of the global dynamics of generic MMRs. 
    
    This paper presents an integrable model that approximates the dynamics of generic MMRs in the {planar} unrestricted planetary three-body problem. 
    This model was used previously in \citet{Hadden2018} to derive a resonance-overlap criterion that predicts the onset of chaos for two massive, eccentric planets.
    An integrable treatment of higher-order resonances provides a key step in extending past criteria for the onset of chaos for nearly circular orbits \citep{Wisdom1980,Deck2013overlap}, based on the overlap of first-order resonances, to the case of eccentric planets for which higher-order resonances play an important role. 
    Here I derive the integrable model for arbitrary-order MMRs and explore some aspects of the secular evolution of resonant planets.
    The model is derived in essentially the same manner as \citet{Delisle2014}'s, utilizing the canonical transformation transformation of \citet{wisdom1986} and \cite{Henrard1986} to concentrate the resonant dynamics in a single ``mode".
    However, I show that this model provides a surprisingly accurate \emph{global} description of the resonant dynamics and need not be restricted to the vicinity of an ACR as \citet{Delisle2014} assume. 

    This paper is organized as follows: I derive the integrable model for resonant motion in Section \ref{sec:analytic_model}. Section \ref{sec:numerical} compares the the integrable model with numerical integrations of resonant planet pairs. Section \ref{sec:secular} explores the secular evolution of resonant planet pairs.
     Finally, I conclude with a summary in Section \ref{sec:summary}.
\section{An Integrable Model for Mean Motion Resonances}
\label{sec:analytic_model}

    In this section I derive an integrable model for the dynamics of two massive planets in or near an MMR of arbitrary order. 
    I consider a pair of planets with masses $m_i$ orbiting in or near a $j$:$j-k$ resonance with $j/(j-k)<2$ around a star of mass $M_*$.  
    In Section \ref{sec:analytic_model:derivation}, I present a Hamiltonian model governing the resonant dynamics of the planet pair.
    In Section \ref{sec:analytic_model:single_dof}, I show that, working to leading order in eccentricity, the resonant dynamics are well-approximated as a one-degree-of-freedom system. 
    In Section \ref{sec:analytic_model:beyond_leading_order}, I show that the one-degree-of-freedom approximation holds even when accounting for terms beyond leading order in eccentricity.

\subsection{Formulation of Resonance Hamiltonian}
\label{sec:analytic_model:derivation}
    The Hamiltonian governing the dynamics of a resonant planet pair, in terms of the canonical {modified Delaunay} variables  \citep[e.g.,][]{MorbidelliBook2002} 
\begin{eqnarray*}
    \Lambda_i &=& \mu_i\sqrt{G(M_*+m_i)a_i}  \\
    \Gamma_i &=& \Lambda_i(1-\sqrt{1-e_i^2})~,
\end{eqnarray*}
    {where $\mu_i=\frac{m_iM_*}{(M_*+m_i)}$,} and their conjugate angle variables $\lambda_i=M_i+\varpi_i$, $\gamma_i=-\varpi_i$ (with $i=1,2$ referring to the inner and outer planet, respectively) is
\begin{equation}
H = H_\text{Kep} + H_\text{res}\label{eq:analytic_model:full_hamiltonian}
\end{equation}
    where
\begin{eqnarray}
H_\text{Kep}= -\frac{G^2}{2}\sum_{i=1}^2 (M_*+m_i)^2\mu_i^3\Lambda_i^{-2} \label{eq:analytic_model:kep_hamiltonian}
\end{eqnarray}
    and
\begin{multline}
H_\text{res}= -\frac{G^2(M_*+m_2)\mu_2^2m_2m_1}{\Lambda_2^2}R_\text{res}(\alpha,e_1,e_2,\varpi_1,\varpi_2,Q)
 \label{eq:analytic_model:res_hamiltonian}
\end{multline}
    where $R_\text{res}$, the resonant disturbing function, is a function of $\alpha=a_1/a_2$, the planets' eccentricities and longitudes of periapse, and $Q=j\lambda_2-(j-k)\lambda_1$.
    The resonant disturbing function is defined as
    \begin{multline}
    R_\text{res}(\alpha,e_1,e_2,\varpi_1,\varpi_2,Q) = \frac{1}{2\pi}\int_0^{2\pi}\frac{a_{2}}{|\vec{r}_2-\vec{r}_1|}d\kappa        -R_\text{sec} \label{eq:analytic_model:Rres}~,
    \end{multline}
    where the integration variable is $\kappa = (\lambda_2-\lambda_1) / k $,  $\vec{r}_1$  and $\vec{r}_2$ are the planets' position vectors,
      and the secular component of the disturbing function, 
\begin{eqnarray*}
    R_\text{sec} = \frac{1}{4\pi^2}\int_{0}^{2\pi}d\lambda_2\int_{0}^{2\pi} d\lambda_1\frac{a_{2}}{|\vec{r}_2-\vec{r}_1|}~,
\end{eqnarray*}
    has been subtracted in order to isolate purely resonant interactions.\footnote{
    The indirect component of the disturbing function, represented by an additional term {$\propto \dot{\vec{r_1}}\cdot \dot{\vec{r_2}}$}, in the integrand of Equation \eqref{eq:analytic_model:Rres} has been omitted as it does not contribute to the averaged disturbing function for the resonances interior to the 2:1 MMR considered in this paper.
    } 
    Orbital elements appearing in $R_\text{res}$ are considered as functions of the canonical \deleted{Poincare} variables in the definition of $H_\text{res}$ (Eq. \ref{eq:analytic_model:res_hamiltonian}).

    Frequently one works with an expansion of $R_\text{res}$ in powers of eccentricity truncated at some maximum order \citep[e.g.][]{Laskar1995}. In Section \ref{sec:analytic_model:single_dof}, I approximate $R_\text{res}$ by the leading-order terms of such an expansion which for a $k$th order resonance are $\propto e^{k}$. Alternatively, Equation \eqref{eq:analytic_model:Rres} can be integrated numerically by evaluating the planets' position vectors, $\vec{r}_1$  and $\vec{r}_2$, as functions of the mean longitudes,  $\lambda_1 = Q/k - j \kappa$ and  $\lambda_2 = Q/k - (j-k) \kappa$, as well as the planets' eccentricities, $e_i$ and longitudes of periapse, $\varpi_i$, in the integrand \citep[e.g.,][]{Schubart1966}.  This approach is taken in Section \ref{sec:analytic_model:beyond_leading_order}.

    The Hamiltonian in Equation \eqref{eq:analytic_model:full_hamiltonian} has four dynamical degrees of freedom. Therefore, three integrals of motion (in addition to the total energy) are required to derive an integrable model for the motion.  I derive three such (approximately) conserved quantities below via a series of canonical transformations. Before proceeding through any canonical transformations, however, I expand the Hamiltonian, Equation \eqref{eq:analytic_model:full_hamiltonian}, about an exactly resonant configuration and re-scale the momenta and energy in order to make the derivation less cumbersome. Defining a reference semi-major axis, $a_{2,0}$, near the outer planet's initial semi-major axis, I choose units so that $\sqrt{G(M_*+m_2)/a_{2,0}^3}=1$. With this reference semi-major axis defined,  the Hamiltonian and all canonical momenta are re-scaled so that 
\begin{equation*}
\{H',\Lambda_i',\Gamma_i'\} =  \frac{2}{(\mu_1+\mu_2)\sqrt{G(M_*+m_2)a_{2,0}}}\{H,\Lambda_i,\Gamma_i\}
\end{equation*}
    are the new Hamiltonian and canonical momenta. I also define {$a_{1,0}=\fracbrac{j-k}{j}^{2/3}\fracbrac{M_*+m_1}{M_*+m_2}^{1/3} a_{2,0}$} as the semi-major axis corresponding to the nominal $j$:$j-k$ interior resonance with a planet at $a_{2,0}$. Next, I transform the canonical momenta conjugate to $\lambda_i$ by adding a constant so that the new momenta are
        \begin{eqnarray}
        \delta\Lambda_i&=&\Lambda_i'-2\frac{\mu_i}{\mu_1+\mu_2}\sqrt{\fracbrac{M_*+m_i}{M_*+m_2}\frac{a_{i,0}}{a_{2,0}}}~.
        \end{eqnarray}
    I assume that the variations on the planets' semi-major axes are small and expand $H'_\text{Kep}$ to second order in $\delta\Lambda_i$ so that the Keplerian component of the Hamiltonian, Equation \eqref{eq:analytic_model:kep_hamiltonian}, becomes
\begin{equation}
H'_\text{Kep}\approx \delta\Lambda_2 +\frac{j}{j-k}\delta\Lambda_1 -\frac{3(\mu_1+\mu_2)}{4}\left(\frac{\delta\Lambda_2^2}{\mu_2}+\frac{\delta\Lambda_1^2}{\mu_1\alpha_0^2}\right)~.\label{eq:analytic_model:kep_hamiltonian_rescaled}
\end{equation}
    {where $\alpha_0=a_{1,0}/a_{2,0}$} and Equation \eqref{eq:analytic_model:res_hamiltonian} becomes  $H'_\text{res} = -2\epsilon R_\text{res}$ where $\epsilon=\frac{m_1\mu_2}{M_*(\mu_1+\mu_2)}$ is of order the planets' mass ratio relative to the star. 
    Finally, I will make the common approximation of ignoring temporal variations of the semi-major axes in the resonant Hamiltonian terms and always evaluate $R_\text{res}$ at {$\alpha = \alpha_0$}. 

    Noting that the planets' mean longitudes only appear in $H'_\text{res}$ in the combination $Q = j\lambda_2+(k-j)\lambda_1$, we can reduce the number of degrees of freedom through the canonical transformation generated by the type-2 generating function
    \begin{multline*}
     F_2(P,{\mathcal K}; \lambda_1,\lambda_2) =  [j\lambda_2 -(j-k)\lambda_1]P +
     \frac{1}{k}(\lambda_2-\lambda_1)({\scK-\scK_0})
    \end{multline*}
    where $\scK_0$ is a constant whose value is chosen to eliminate terms linear in $P$ from the transformed Hamiltonian. The canonical transformation yields the conserved quantity ${\mathcal K}=\scK_0-(j\delta\Lambda_1+(j-k)\delta\Lambda_2)$, since its conjugate angle $\kappa=(\lambda_2-\lambda_1)/k$ does not appear in the Hamiltonian. 
    After dropping constant terms  the transformed Hamiltonian becomes
    \begin{eqnarray}
    H''(P, Q,\Gamma_i,\gamma_i) = -\frac{1}{2}A P^2 + H'_\text{res}(Q,\Gamma_i,\gamma_i)
    \end{eqnarray}
    where 
    \begin{eqnarray}
        A = \frac{3j(\mu_1+\mu_2)}{2}\left(\frac{j}{\mu_2}+\frac{(j-k)}{\mu_1\sqrt{\alpha_0}}\right)~.
    \end{eqnarray}
    The distance of the planets from exact commensurability is related to the new canonical momentum, $P$, by
    \begin{eqnarray}
    \frac{(j-k)}{j}\frac{P_2}{P_1}-1
    &\approx \frac{3(\mu_1+\mu_2)}{2\mu_2}\delta\Lambda_2-\frac{3(\mu_1+\mu_2)}{2\mu_1\sqrt{\alpha_0}}\delta\Lambda_1 \nonumber\\
    &=\frac{3(\mu_1+\mu_2)}{2}\left[\frac{j}{\mu_2} + \frac{j-k}{\mu_1\sqrt{\alpha_0}}\right] P\label{eq:analytic_model:period_ratio}~.
    \end{eqnarray}

\subsection{Reduction to one degree of freedom}
\label{sec:analytic_model:single_dof}
\begin{figure}
    \centering
    \includegraphics[width=\columnwidth]{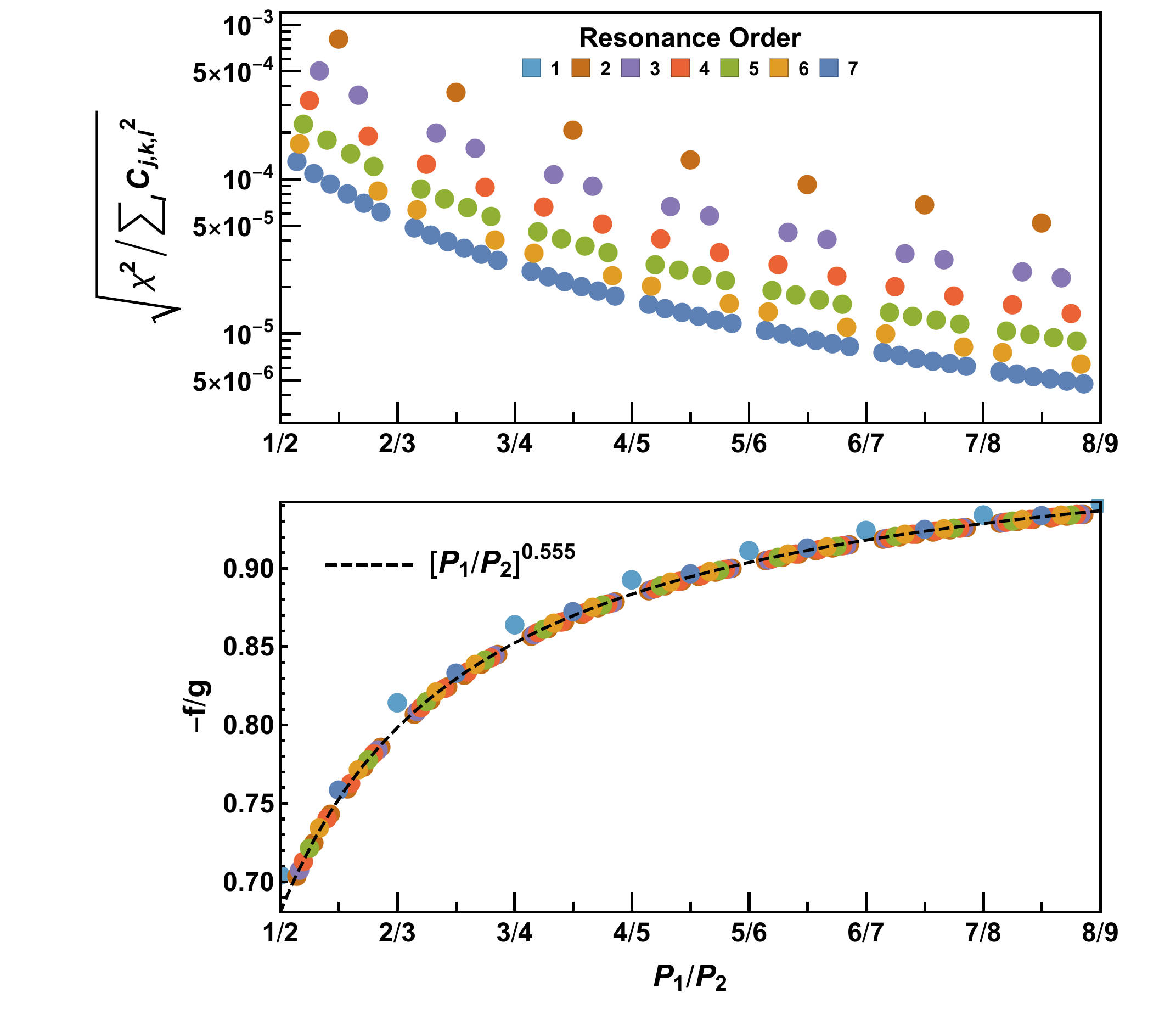}
    \caption{The top panel shows $\left(\chi^2 / \sum_{l=0}^{k}|C_{j,k,l}|^2\right)^{1/2}$, a measure of the fractional error of Equation \eqref{eq:analytic_model:magic_transform}, where $\chi^2$ is defined in Equation \eqref{eq:analytic_model:chi_sq_def}, versus resonant period ratio for resonances up to 7th order between the 2:1 and 9:8 MMRs. The bottom panel plots the ratio $-f/g$ for the same resonances plotted in the top panel, along with a power-law fit $f/g\approx -(P_1/P_2)^{0.55}$ shown as a dashed line.}
    \label{fig:analytic_model:fg_plots}
\end{figure}

    Assuming the planets' eccentricities are small, the resonant dynamics can be approximated by taking the leading-order terms of an expansion of $R_\text{res}$ in powers of eccentricity. To leading order, 
    \begin{eqnarray}
    R_\text{res}&\approx& \sum_{l=0}^{k}C_{j,k,l}e_1^{l}e_2^{k-l}\cos[Q-l\varpi_1-(k-l)\varpi_2]\nonumber\\
    &=&\frac{\exp[iQ]}{2}\sum_{l=0}^{k}C_{j,k,l}z_1^{*l}z_2^{*(k-l)} + c.c.
    \label{eq:analytic_model:single_dof:DFexpansion}
    \end{eqnarray}
    where the $C_{j,k,l}$ are combinations of Laplace coefficients and their derivatives \citep[e.g.,][]{MD1999ssd,MorbidelliBook2002}, $z^*_i=e_i\exp(-i\varpi_i)$ the complex conjugates of the planets' complex eccentricities, and `$c.c.$' denotes the complex conjugate of the preceding term.
     Thus, to leading order in eccentricity, $H'_\text{res}(Q,\Gamma_i,\gamma_i)$, contains  $k+1$ distinct ``sub-resonances" with resonant angles 
\begin{equation*}
\theta_{res,l}=Q-l\varpi_1-(k-l)\varpi_2
\end{equation*} 
    for $l=0,1,...,k$ so it seems that there is no simple canonical transformation that can reduce the Hamiltonian to  dependence on a single resonant angle. This apparent difficulty can be surmounted by exploiting the surprising fact that, to an excellent approximation, the sum appearing in Equation \eqref{eq:analytic_model:single_dof:DFexpansion} can be re-written as a single linear combination of $z_1^*$ and $z_2^*$ raised to the $k$th power. In other words, there exist coefficients $f$ and $g$ such that  
\begin{equation}
\sum_{l=0}^{k}C_{j,k,l}z_1^{*l}z_2^{*k-l}\approx \left(fz_1^*+gz_2^*\right)^{k}~.
\label{eq:analytic_model:magic_transform}
\end{equation}
 While no choice of $f$ and $g$ can make Equation \eqref{eq:analytic_model:magic_transform} exact (except in the trivial case $k=1$), the approximation is an excellent one. Equation \eqref{eq:analytic_model:magic_transform} presents an over-determined system of equations for the  $k+1$  coefficients $C_{j,k,l}$ in terms of two unknowns, $f$ and $g$. In this paper I determine values for $f$ and $g$ via least-squares regression by minimizing
\begin{eqnarray}
    \chi^2(f,g) = \sum_{l=0}^{k} \left[C_{j,k,l}-\binom{k}{l}f^{l}g^{k-l}\right]^{2}~\label{eq:analytic_model:chi_sq_def}
\end{eqnarray}
    which is the sum of the squared differences of monomial coefficients between the left- and right-hand side of Equation \eqref{eq:analytic_model:magic_transform}.
    Figure \ref{fig:analytic_model:fg_plots} plots a measure of the fractional error in the approximation Equation \eqref{eq:analytic_model:magic_transform} for resonances  between the 2:1 and 9:8 MMRs up to order $k=7$, showing that the agreement is excellent and improves with both closer planet spacings and higher resonance order.
    The dynamical origins of the approximate equality expressed by Equation \eqref{eq:analytic_model:magic_transform} are discussed in Section \ref{sec:why}.

    In light of Equation \eqref{eq:analytic_model:magic_transform}, I define  
\begin{equation}
    \begin{pmatrix}
        Ze^{iz}\\
        We^{iw}
    \end{pmatrix}
        = 
    \frac{1}{\sqrt{f^2+g^2}}\begin{pmatrix}
        f & g \\
        -g & f
    \end{pmatrix} 
    \cdot 
    \begin{pmatrix}
        z_1\\
        z_2
    \end{pmatrix}
\label{eq:analytic_model:Zdef}
\end{equation}
     so that Equation \eqref{eq:analytic_model:single_dof:DFexpansion} approximately reduces to 
     \begin{eqnarray}
        R_\text{res}\approx (f^2+g^2)^{k/2}Z^k\cos[Q-kz]
        \label{eq:analytic_model:single_dof:Rres_approx}
     \end{eqnarray}
     and is independent of $W$ and $w$. For all MMRs interior to the 2:1 MMR, $f \approx -g$ so that ${Z}e^{iz}$ is roughly the vector difference of the planet pair's eccentricity vectors represented in the complex plane. The ratio $f/g$ is more precisely fit as the power-law $f/g\approx -(P_1/P_2)^{0.55}$, which is plotted in the bottom panel of Figure \ref{fig:analytic_model:fg_plots}.
     This approximation does \emph{not} apply to the 2:1 MMR itself due to the presence of indirect terms in the disturbing function \citep[e.g.,][]{MD1999ssd}.

    Now that $R_\text{res}$ is expressed via Equation \eqref{eq:analytic_model:single_dof:Rres_approx} in terms of a  single cosine term, we are in a position to derive an integrable Hamiltonian model. First, we need to express $Z$ and $z$ in terms of the canonical variables. 
    This is accomplished by first noting that, to leading order in eccentricity, $z_1^* \approx {\fracbrac{\mu_1+\mu_2}{\mu_1\sqrt{\alpha_0}}}^{1/2}\sqrt{\Gamma'_1}e^{i\gamma_1}$ and $z_2^* \approx {\fracbrac{\mu_1+\mu_2}{\mu_2}}^{1/2} \sqrt{\Gamma'_2}e^{i\gamma_2}$.
    Then, the canonical rotation transformation 
\begin{equation}
\begin{pmatrix}
\sqrt{\Phi}e^{i\phi} \\ \sqrt{\Psi}e^{i\psi}
\end{pmatrix}=
\frac{1}{\sqrt{\tilde{f}^2+\tilde{g}^2}}\begin{pmatrix}
\tilde{f} & \tilde{g} \\
- \tilde{g} & \tilde{f}
\end{pmatrix}\begin{pmatrix}
\sqrt{\Gamma'_1}e^{i\gamma_1} \\ \sqrt{\Gamma'_2}e^{i\gamma_2}
\end{pmatrix}
\label{eq:analytic_model:PhiPsi_define}
\end{equation}
where 
\begin{eqnarray*}
\tilde{f} \equiv \sqrt{\frac{\mu_1+\mu_2}{\mu_1\sqrt{\alpha_0}}}f&~\text{and}~& \tilde{g} \equiv \sqrt{\frac{\mu_1+\mu_2}{\mu_2}}g
\end{eqnarray*}
   defines new conjugate momentum-coordinate pairs $(\Phi,\phi)$ and $(\Psi,\psi)$ \citep{wisdom1986,Henrard1986,BatyginMorbidelli2013analytic,Delisle2014}. We now have $\sqrt{\Phi}e^{i\phi} \propto Ze^{-iz}$ and the resonant component of the Hamiltonian becomes
\begin{equation}
    H'_\text{res}\approx -2\epsilon(\tilde{f}^2+\tilde{g}^2)^{k/2}\Phi^{k/2}\cos[ Q+k\phi ]~.%
\label{eq:analytic_model:hres_tranformed}
\end{equation}
   The Hamiltonian is independent of both $\Psi$ and $\psi$ so the complex quantity
\begin{eqnarray}
    \sqrt{\Psi}e^{i\psi} 
    &=& \frac{\mu_2fz_2-\mu_1 g\alpha_0^{1/2} z_1}{\sqrt{(\mu_1\alpha_0^{1/2}g^2+\mu_2f^2)(\mu_1+\mu_2)}}
    \label{eq:analytic_model:Zcom}
\end{eqnarray}
  is approximately conserved by the resonant dynamics.\footnote{
  {The reader is cautioned that, while $\sqrt{\Phi}e^{i\phi} \propto Ze^{-iz}$, it is \emph{not} the case that $\sqrt{\Psi}e^{i\psi} \propto We^{-iw}$. Instead, $\sqrt{\Psi}e^{i\psi}$ is given by a linear combination of $Ze^{-iz}$ and $We^{-iw}$.
  }} 
  In the limit that one planet's mass vanishes, Equation \eqref{eq:analytic_model:Zcom} implies the conservation of the massive planet's eccentricity.

    It is now straightforward to reduce the system to a single degree of freedom using the canonical transformation generated by 
\begin{equation}
F_2(J,J^*;Q,\phi) = \frac{1}{k}(Q+k\phi)J-\frac{1}{k}(Q+k\psi) J^*+\psi {\cal A}
\label{eq:analytic_model:f2-transform}
\end{equation}
yielding $J = \Phi$ and $\theta = \frac{1}{k}Q+\phi$ as a new canonical momentum-coordinate pair along with $J^*=\Phi-k P$ and ${\cal A}=\Psi+\Phi-k P$ as conserved quantities with conjugate cyclic coordinates $\theta^*=-\frac{1}{k}Q-\psi$ and $\psi$, respectively.\footnote{
${\cal A}$ is strictly conserved by the full resonant Hamiltonian, Equation \eqref{eq:analytic_model:full_hamiltonian} because the D'Alembert property of the disturbing function ensures that its conjugate variable, $\psi$, does not appear explicitly in the Hamiltonian. This is most readily seen by considering Equation \eqref{eq:analytic_model:Zcom} in the limit where one of the masses vanishes: $\psi$ is then equal to the longitude of perihelion of the massive planet which cannot appear as an isolated cosine argument in a disturbing function expansion \citep[e.g.,][]{MD1999ssd}. Therefore, unlike ${J^*}$, the conservation of ${\cal A}$ does not rely on the accuracy of the approximation expressed in Equation \eqref{eq:analytic_model:magic_transform}.
} The transformed Hamiltonian is
\begin{eqnarray}
    {\cal H}(J,\theta;J^*)=-\frac{1}{2k^2}A(J-J^*)^2-{\tilde\epsilon}J^{k/2}\cos(k\theta)~,\label{eq:analytic_model:final_hamiltonian}
\end{eqnarray}
    where $\tilde{\epsilon} = 2(\tilde{f}^2+\tilde{g}^2)^{k/2}\epsilon$. Finally, Equation \eqref{eq:analytic_model:final_hamiltonian} gives an integrable one degree-of-freedom Hamiltonian for the resonant dynamics. 
    Hamiltonians of the form given in Equation \eqref{eq:analytic_model:final_hamiltonian} have been studied extensively in the literature as models for resonant motion \citep[e.g.,][]{HenrardLemaitre1983,Lemaitre1984}.
    The dynamics are approximately pendulum-like, with $J$ oscillating about an equilibrium value of $\approx J^*$.

    After the series of canonical transformations it is instructive to express the canonical variables $J$ and $\theta$ in terms of familiar orbital elements. Working back through the transformations, we have 
\begin{eqnarray}
    J  &=& \frac{f^2e_1^2+g^2e_2^2 + 2 f g e_1e_2\cos(\varpi_2-\varpi_1)}{(\tilde{f}^2+\tilde{g}^2)}\nonumber\\
        &=& \frac{f^2+g^2}{\tilde{f}^2+\tilde{g}^2}Z^2  \nonumber\\
    k\theta &=& j\lambda_2 - (j-k)\lambda_1 -kz~.
    \label{eq:analytic_model:J_theta_def}
\end{eqnarray}
    Additionally, using Equation \eqref{eq:analytic_model:period_ratio} and $(J-J^*)/k=P$,  the dynamical variables are related to the planets' instantaneous period ratio by
\begin{equation}
        \frac{j-k}{j}\frac{P_2}{P_1}-1=\frac{3(\mu_1+\mu_2)\left(j\mu_1\sqrt{\alpha_0}+(j-k)\mu_2\right)}{2k\sqrt{\alpha_0}\mu_1\mu_2}(J-J^*)~.
        \label{eq:analytic_model:JJstar_to_Delta}
\end{equation}

\subsection{Beyond the Leading Order Approximation}
\label{sec:analytic_model:beyond_leading_order}
    Hamiltonian \eqref{eq:analytic_model:final_hamiltonian} was derived by considering only the lowest-order terms from the expansion of the $R_\text{res}$, Equation $\eqref{eq:analytic_model:Rres}$, in powers of the planets' eccentricities. The derivation took advantage of a surprising near-symmetry exhibited by these lowest-order terms, expressed in Equation \eqref{eq:analytic_model:magic_transform}, that leads to the (approximate) conservation of $\Psi$. In fact, this near-symmetry persists even when higher order resonant terms are included in the Hamiltonian.
    
    {
    The resonant disturbing function can be computed without using a truncated expansion in eccentricities by evaluating Equation \eqref{eq:analytic_model:Rres} via numerical quadrature. 
    In the following, I evaluate $R_\text{res}$ and its derivatives numerically using a Gauss-Legendre quadrature rule. 
    I utilize the \texttt{exoplanet} package's \citep{exoplanet:exoplanet} Kepler solver, based on algorithms by \citet{Nijenhuis1991} and \citet{Markley1995}, in order to compute planets' position vectors as functions of the canonical variables. Code for evaluating $R_\text{res}$ by numerical quadrature is available online at \website.
    }
    
    The full resonant disturbing function can be represented as the cosine series {(see Appendix \ref{sec:appendix:fourier})}
\begin{multline}
    R_\text{res} = \sum_{n=1}^{\infty}\sum_{l=-\infty}^{\infty}a_{n,l}(Z,W) \cos[nQ - (nk-l)z-l w]\label{eq:beyond_leading_order:Rfull_fourier}
\end{multline}
    where, to leading order in eccentricities, the amplitudes $a_{n,l}(Z,W)\propto Z^{|nk-l|}W^{|l|}$  so that amplitudes decrease with increasing $n$ and the terms with $0\le l\le nk$ are generally the most important for a given $n$ \deleted{(see Appendix \ref{sec:appendix:fourier})}. {
    (The sum of the exponents of $Z$ and $W$ is equal to $nk$ for $0\le l\le nk$ while it is at least $nk+2$ for $l<0$ or $l>nk$.)}
    The approximation for $R_\text{res}$ used in Section \ref{sec:analytic_model:single_dof} to derive an integrable Hamiltonian (Equation \ref{eq:analytic_model:final_hamiltonian}) included only the $(n,l)=(1,0)$ term from the sum in Equation \eqref{eq:beyond_leading_order:Rfull_fourier} 
    and furthermore approximates this term, $a_{1,0}(Z,W)$, by its leading-order piece $\propto Z^{k}$.
    
\begin{figure}
    \centering
    \includegraphics[width=\columnwidth]{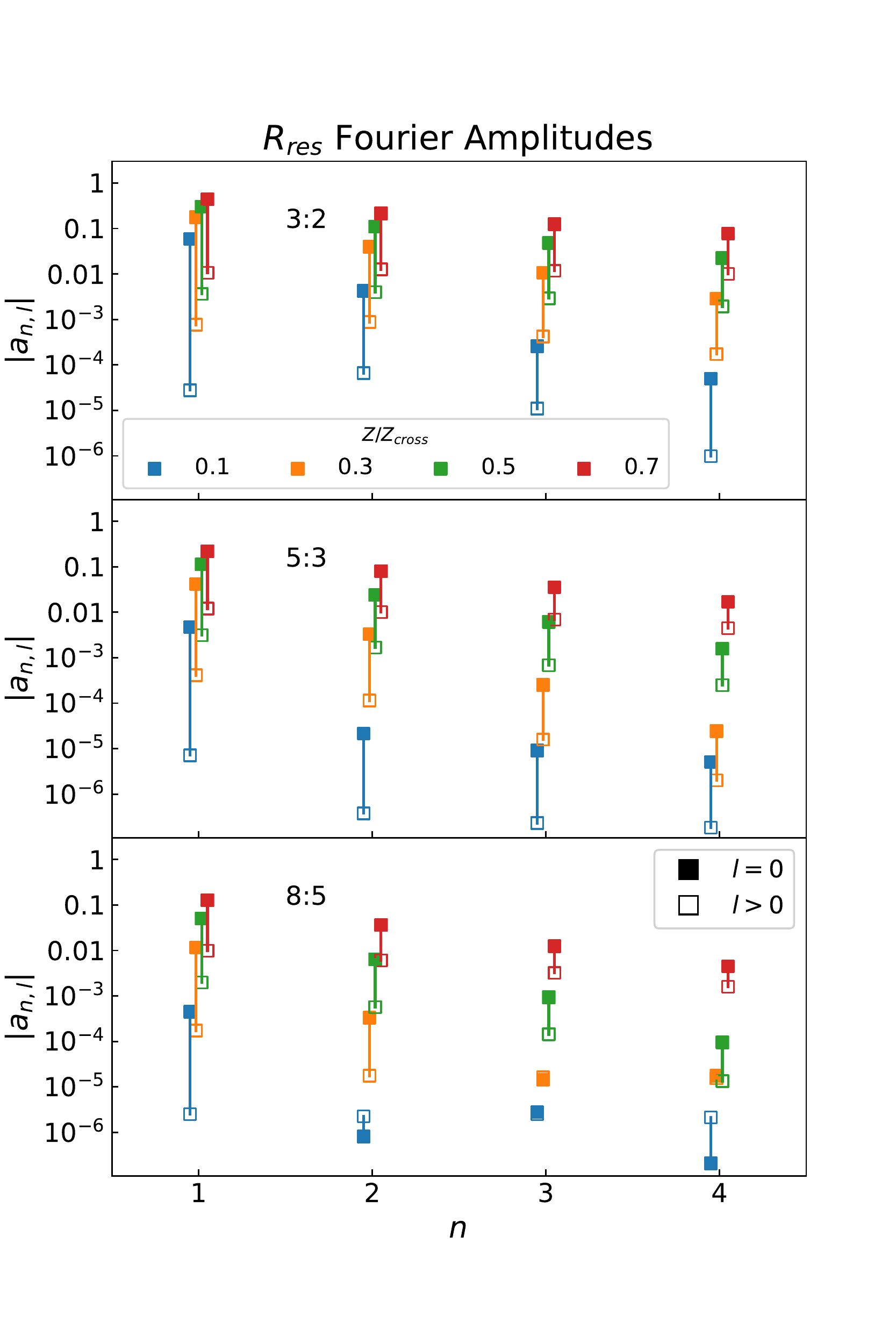}
    \caption{
    Amplitudes, $a_{n,l}$, of the cosine series for the resonant disturbing function $R_\text{res}$ (Equation \ref{eq:beyond_leading_order:Rfull_fourier})  are plotted versus $n$ for different resonances and eccentricities. 
    {
    Amplitudes $a_{n,l}$ are determined from an FFT after evaluating $R_\text{res}$ on a $128\times128$ grid in $z$ and $w$ (with $Q=0$ fixed). The $W=Z$
    }
    Solid squares show the coefficient $a_{n,0}$ while empty squares show the next-largest $a_{n,l}$ with $0<l\le n k$. 
    \deleted{Coefficients are computed from an FFT of a $128\times128$ grid of $R_\text{res}$ values in $z$ and $w$ computed numerically with $W=Z$ and $Q=0$.}
    In each panel, $Z$s are set to a constant fraction of $Z_\text{cross}$, the value at which the planets' orbits first intersect.}
    \label{fig:beyond_leading_order:res_fourier_amps}
\end{figure}

    If, more generally, the resonant disturbing function is approximated as
    \begin{eqnarray}
    R_\text{res}\approx \sum_{n=1}^{\infty}a_{n,0}(Z,W)\cos[n(Q-kz)]~\label{eq:analytic_model:Rres_full_approx},
    \end{eqnarray}
    then it still depends only on a single resonant angle, $k\theta = Q-kz$ and $\Psi$ will remain approximately conserved by the resonant dynamics. 
    Figure \ref{fig:beyond_leading_order:res_fourier_amps} shows that Equation \eqref{eq:analytic_model:Rres_full_approx} is in fact a remarkably good approximation of the resonant disturbing function even in cases where Equation \eqref{eq:analytic_model:single_dof:Rres_approx} is not.
    The figure shows the magnitudes of Fourier amplitudes, $a_{n,l}$, for three different resonances and a range of eccentricities. As the planets approach orbit-crossing, high-order Fourier coefficients decay more slowly and terms in Equation \eqref{eq:beyond_leading_order:Rfull_fourier} with $n>1$ become increasingly important. However, $a_{n,0}$ generally remains larger than the next-largest $a_{n,l}$ by roughly an order of magnitude or more at each $n$. In all but one case, $a_{n,0}>a_{1,l}$ with $l>0$ up to at least $n=4$. 
    
    The sum in Equation \eqref{eq:analytic_model:Rres_full_approx} can be computed by simply evaluating $R_\text{res}$ with $W=0$ if fractional corrections of order ${\cal O}(W^2)$ to the $a_{n,0}$ amplitudes are ignored. Accordingly, the Hamiltonian governing the resonant dynamics can be approximated by the integrable Hamiltonian
\begin{multline}
    {\cal H}(J,\theta; J^*)\approx
    -\frac{1}{2k^2}A(J-J^*)^2+H'_\text{res}(J,\theta)
    \label{eq:beyond_leading_order:Hfull}    
\end{multline}
where 
\begin{eqnarray*}
H'_\text{res}(J,\theta) = -2\epsilon  R_{\text{res},W=0}~
\end{eqnarray*}
is computed by numerically integrating Equation \eqref{eq:analytic_model:Rres}.
This Hamiltonian is compared with $N$-body results below in Section \ref{sec:numerical}. 

\subsection{Hill's problem and the origin of the reducing transformation}
\label{sec:why}
    Equation \eqref{eq:analytic_model:magic_transform} expresses an apparently coincidental relationship between disturbing function coefficients that yields an approximately-conserved quantity, $\Psi$, and makes the derivation of the integrable approximation of the resonant dynamics possible. 
    This naturally leads to the question:  what near-symmetry of the underlying dynamics is responsible for the existence of this approximately-conserved quantity?
    Hill's equations \citep{Hill1878} offer some insight into the answer.  
    
    As \citet{Henon1986} note, Hill's equations are not merely a limiting case of the circular restricted three-body problem as they do not place any restrictions on the eccentricities of the orbits under consideration nor do they require that one body's mass vanish. Instead, Hill's approximation only requires two bodies' (i.e., the planets) masses be small compared to the primary (i.e., the star).  Under Hill's approximation, the small bodies' motion is separated into the center-of-mass motion, which follows a Keplerian orbit with complex eccentricity $(\mu_1 z_1+ \mu_2 z_2)/(\mu_1+\mu_2)$, and the relative motion of the bodies, which evolves according to Hill's equations.   In particular, the evolution of the planets' orbits under Hill's approximation can be expressed in terms of a fictitious \emph{relative} orbit with orbital elements that include a \emph{relative} complex eccentricity, $z_2-z_1$.

    Building on the work of \citet{Henon1986}, \citet{Duncan1989} and \citet{Namouni1996} derive area-preserving maps that approximate the dynamics of a pair of planets by ``stitching together" successive close encounters approximated by Hill's equations. These area-preserving maps successfully reproduce resonant phase-space structure seen in $N$-body integrations (though, as \citet{Namouni1996} demonstrate, the mapping must include sufficiently high-order terms in eccentricity to successfully capture the phase-space structure of high-order resonances for eccentric planets). Therefore, to the degree that resonant dynamics are accurately approximated by these `encounter maps' the resonant dynamics should, just like Hill's equations, only depend on the \emph{relative} eccentricity, $z_2-z_1$.
    
    In the limit $P_2/P_1\rightarrow 1$, the coefficients $f/g \rightarrow -1$ so that $Ze^{iz}\rightarrow \frac{1}{\sqrt{2}}(z_2-z_1)$ and the conserved quantity $\sqrt{\Psi}e^{i\psi}\propto (\mu_1 z_1+ \mu_2 z_2)/(\mu_1+\mu_2)$ (Equation \ref{eq:analytic_model:Zcom}). Therefore, in the close-spacing limit, the integrable model depends only on the complex relative eccentricity and conserves the complex eccentricity of the center-of-mass orbit like Hill's equations.
    
\section{Comparison with Numerical Integrations}
\label{sec:numerical}
    Here I compare the integrable resonance models derived in Sections \ref{sec:analytic_model}  with  $N$-body integrations.  
    All $N$-body integrations throughout this paper are done with the \texttt{WHFast} integrator \citep{RTwhfast2015} based on the symplectic mapping algorithm of \citet{wisdom_holman1991} and implemented in the \texttt{REBOUND} code \citep{RL12}. 
    Integration step sizes are set to 1/30th of the perihelion passage timescale of the inner planet, defined as $T_{p}=2\pi/\dot{f}_p$ where $\dot{f}_p$ is the rate of change of the true anomaly at pericenter \citep{wisdom2015}. 
    While the analytic model applies to all MMRs of any order interior to the 2:1 MMR, the numerical examples throughout the rest of the paper will focus on the 3:2, 5:3, and 8:5 MMRs.\footnote{{Technically, the canonical variables appearing in the resonance Hamiltonian introduced in Equation \eqref{eq:analytic_model:full_hamiltonian} differ from standard canonical variables of the full $N$-body problem by a near-identity transformation \cite[see, e.g., ][ for an explicit construction of this transformation to first order in planet masses and eccentricities]{Deck2013overlap}. I do not make any corrections for this difference when comparing $N$-body integration with the analytic model as they are negligible in the cases studied here. However, this effect becomes more significant as MMRs' orbital separations become closer.}}

    I show in Appendix \ref{sec:appendix:fourier} that all resonances of a given order exhibit essentially the same functional dependence $Z$ when scaled by $Z_\text{cross}$, the value at which the planets' orbits first intersect, so these resonances serve as representative examples for all first-, second-, and third order resonances. The value of $Z_\text{cross}$ is determined for fixed values of $W$, $w$, and $z$ by solving for the value of $Z$ for which the planets' complex eccentricities satisfy
    \begin{equation*}
        \alpha^2  (1-|z_1|^2) + (1-|z_2|^2) - \alpha  ( 2 -  z_1 z^*_2 - z^*_1 z_2)=0~
    \end{equation*}
   \citep{Kholshevnikov1999}. For closely spaced planets, approximating $f/g \approx -1$ and writing $\alpha = 1-\delta$ gives  
   $Z_\text{cross}\approx \frac{1}{\sqrt{2}}\delta(1+
   \frac{W}{\sqrt{2}}\cos(z-w))$ to first order in $\delta$ and $W$.
    
\begin{figure*}
    \centering
    \includegraphics[width=\textwidth]{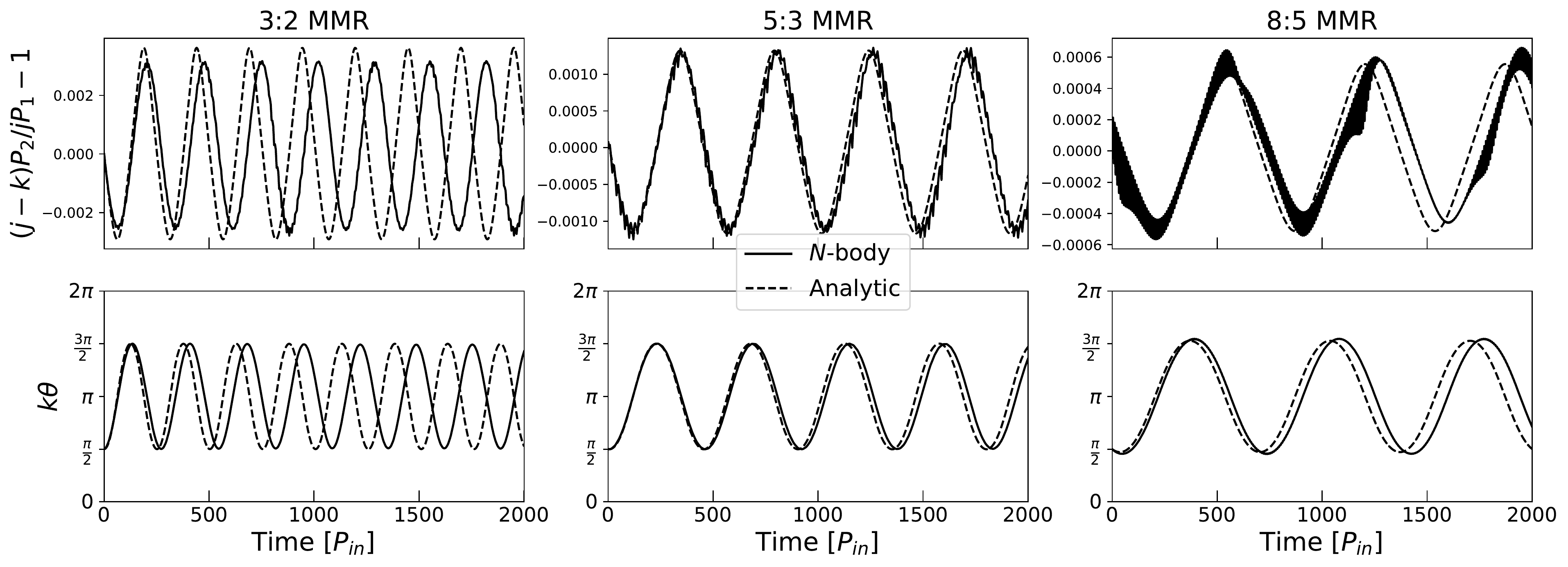}
    \caption{Resonant motion of two massive planets in three different resonances. The top panels show the deviation of the planets' osculating period ratio from exact resonance and the bottom panels show librations of the resonant angles $k\theta$ (Equation \eqref{eq:analytic_model:J_theta_def}. Solid lines show the results of $N$-body integrations and dashed lines show the results of integrating the equations of motion derived from the Hamiltonian \eqref{eq:analytic_model:final_hamiltonian}. Both planets have a mass of $m_1=m_2=10^{-5}M_*$ and are started at the nominal location of the resonance so that $a_{2}=\fracbrac{j}{j-k}^{2/3}a_{1}$. The eccentricity of the inner planet is $e_{1}=0.05$ in each case.  The planets' orbits are initially aligned ($\varpi_1=\varpi_2$) and the eccentricity of the outer planet is set so that $Z$ is 15\% of $Z_\text{cross}$, the value at which the planets' orbits cross. The eccentricity of  the outer planet is set to  $e_2=0.082,0.089,$ and $0.086$ for the 3:2, 5:3, and 8:5 MMR, respectively. The initial mean longitude of the other planet is chosen so that $k\theta = \pi/2$.}
    \label{fig:numerical:tri_plot}
\end{figure*}

    Figure \ref{fig:numerical:tri_plot} shows some examples of the resonant motion of pairs of massive planets in different resonances, comparing the simple model derived to leading order in eccentricity in Section \ref{sec:analytic_model:single_dof} with $N$-body integrations. The analytic model shows good general agreement with the $N$-body results; the most significant differences being slight deviations in the predicted libration frequencies, especially in the case of the first-order 3:2 MMR. 
    {
    In this case, the disagreement is primarily due to truncating $R_\text{res}$ at first order in $Z$; the disagreement is greatly reduced by including the additional term $\propto Z^2\cos[2(Q-z)]$ from $R_\text{res}$ in integrations of the analytic model.
    }

\begin{figure}
    \centering
    \includegraphics[width=\columnwidth]{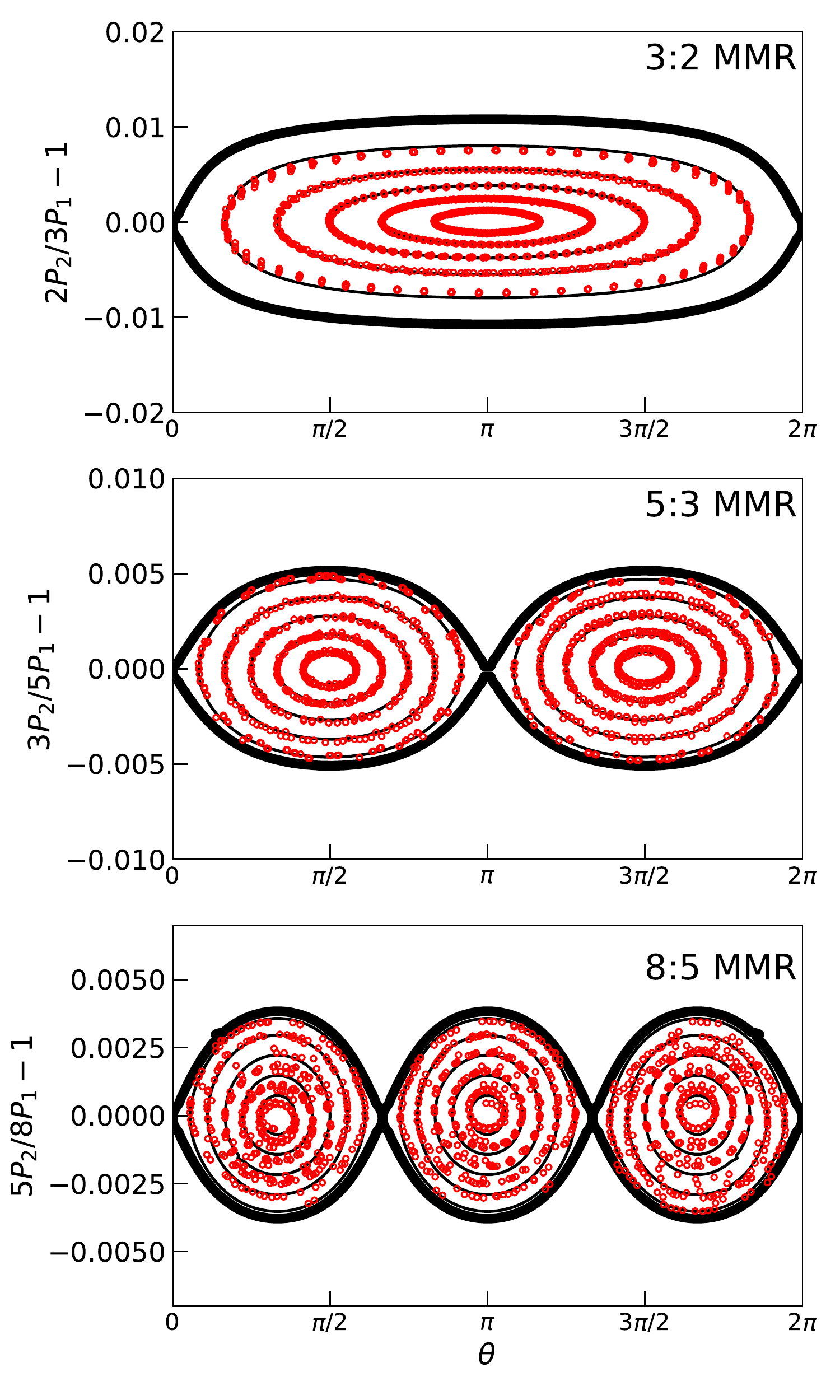}
    \caption{
    A comparison between Hamiltonian \eqref{eq:beyond_leading_order:Hfull} 
    and $N$-body integrations of resonant pairs of planets with masses $m_1=m_2=10^{-5}M_*$. 
    Black lines show contour levels of the Hamiltonian while red points are taken from $N$-body integrations spanning 500 orbital periods of the outer planet. Initial conditions are chosen so that $a_1=\fracbrac{j-k}{j}^{2/3}a_2$, $(W,w,Z,z)=(0.1,0,0.5Z_\text{cross},0)$ and $\lambda_1=0$. The initial value of $\lambda_2$ is varied to create different initial values of $\theta$ to match the thin plotted Hamiltonian contours at $(j-k)P_2/jP_1-1=0$. The separatrix of Hamiltonian \eqref{eq:beyond_leading_order:Hfull} is also plotted as a thick black contour.}
    \label{fig:numerical:Hres_contours}
\end{figure}

    Figure \ref{fig:numerical:Hres_contours} compares the numerically-computed Hamiltonian model, Equation \eqref{eq:beyond_leading_order:Hfull}, with $N$-body integrations. The Hamiltonian model shows excellent agreement with the $N$-body integrations. The $N$-body integrations were initialized with $W=0.1$ while Hamiltonian \eqref{eq:beyond_leading_order:Hfull} is computed assuming $W=0$, confirming that the resonant dynamics are nearly independent of $W$. A moderately large value of $Z\approx 0.5 Z_\text{cross}$ was chosen in Figure \ref{fig:numerical:Hres_contours}  in order to illustrate a regime where contours of the numerically-averaged Hamiltonian  differ significantly from those of the simple Hamiltonian \eqref{eq:analytic_model:final_hamiltonian}.
    
\begin{figure*}
    \centering
    \includegraphics[width=\textwidth]{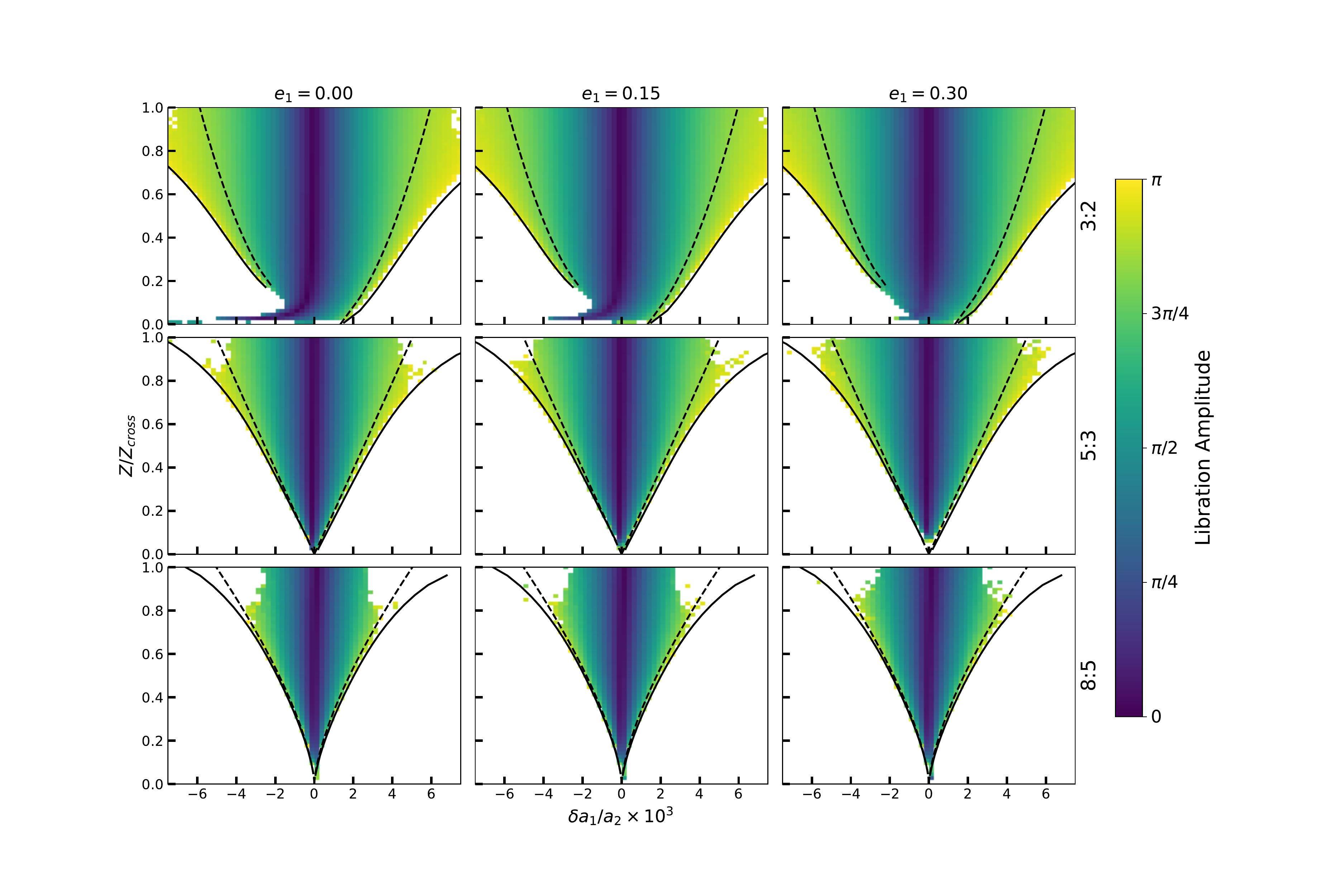}
    \caption{
    Maps of resonant libration amplitudes for different resonances and eccentricities from $N$-body simulations. 
    Each column corresponds to a different initial eccentricity of the inner planet and each row shows a different MMR.
    In each panel, the libration amplitude of the resonant angle, $k\theta$ (Eq \ref{eq:analytic_model:J_theta_def}), is recorded from $N$-body simulations on a grid of initial conditions and indicated by the color scale. Grid points where the resonant angle circulates or alternate between libration and circulation are not colored.     
    Integrations are run for 250 orbital periods of the outer planet. Initially, $\varpi_2-\varpi_1=\pi$ and $\theta=\pi/k$. Separatrices for the Hamiltonians \eqref{eq:analytic_model:final_hamiltonian} and \eqref{eq:beyond_leading_order:Hfull} are shown by dashed and solid lines, respectively.
    The similarity of results in each row serves to illustrate that, to excellent approximation, the resonant dynamics depend on the planets' eccentricities only through the combination $Z$.
}
    \label{fig:numerical:res_grid}
\end{figure*}

    The analytic model predicts that systems with different individual eccentricities but the same combined eccentricity, $Z$, will have the same resonant dynamics. Figure \ref{fig:numerical:res_grid} illustrates this by comparing a series of $N$-body integrations at different resonances that are initialized with different individual planet eccentricities but equivalent $Z$s. Each panel shows a map of the resonant angle libration amplitude, measured from $N$-body integrations.  Different resonances are plotted in each row, and each column corresponds to a different initial eccentricity of the inner planet. The libration amplitude maps for a given resonance are nearly indistinguishable over the wide range of inner planet eccentricity spanning $e_{1}=0$  to $0.3$.

    Resonance widths predicted with the simple leading-order model, Equation \eqref{eq:analytic_model:final_hamiltonian}, as well as the numerically-averaged model, Equation \eqref{eq:beyond_leading_order:Hfull} are shown in each panel. Separatrices are computed as follows: for a fixed value of $J^*$, the unstable fixed point of ${\cal H}(J,\theta; J^*)$ is located at $\theta=0$ by numerically solving for the value of $J$ that maximizes ${\cal H}(J,0; J^*)$. The maximum, $E_{sx}(J^*)$, determines the energy of the separatrix trajectory.
    The maximal width of the separatrix is then determined by solving ${\cal H}(J,\pi/k)=E_{sx}(J^*)$ for $J$. The values of $J$ and $J^*$ are then converted to $Z$ and $\delta a_1 /a_2$ using Equations \eqref{eq:analytic_model:J_theta_def} and \eqref{eq:analytic_model:JJstar_to_Delta}.
    The most significant differences between the leading-order model and the full numerically-averaged model occurs for the 3:2 first-order MMR.

%

\section{Secular Evolution}
\label{sec:secular}
    In this section I consider resonant dynamics with a more complete Hamiltonian model that includes secular terms. 
    {    
    Secular dynamics inside MMRs have been examined by numerous previous studies, usually with applications to small solar system bodies \citep[e.g.,][]{Kozai1985,Wisdom1985,Yoshikawa1989,MorbidelliMoons1993,MoonsMorbidelli1995,Gallardo2012,Saillenfest2016}, 
    though extra-solar contexts have also been considered \citep[e.g.,][]{Beust1996,BatyginMorbidelli2013analytic,Pichierri2017}.
    The treatment presented here is similar to \citet{BatyginMorbidelli2013analytic}, though here I generalize to MMRs of arbitrary order.}\footnote{
    \citet{BatyginMorbidelli2013analytic} also truncate the resonant disturbing function at first eccentricity whereas here I evaluate the resonant disturbing by numerical quadrature. This can yield some differences in the quantitative predictions of the model as already seen, for example, in Figure \ref{fig:numerical:res_grid}.
    }

    In Appendix \ref{sec:appendix:secular_derivation} I show that, to leading order in eccentricity, secular interactions between a planet pair are governed by the Hamiltonian
\begin{multline}
H'_\text{sec}=-\epsilon\left[a_sJ - b_s J^* +c_s\sqrt{J({\cal A}-J^*)}\cos(\theta+\theta^*)\right]
\label{eq:secular:sec_ham}
\end{multline}
    where coefficients $a_s, b_s$ and $c_s$ depend on $\alpha$ and the planets' mass ratio, $m_1/m_2$. Explicit expressions are given in Appendix \ref{sec:appendix:secular_derivation}.
    
    The inclusion of secular terms has introduced an explicit dependence on $\theta^*$ to the Hamiltonian so that $J^*$  is no longer a conserved quantity. However, ${\cal A}=\Psi+\Phi - kP$ is still conserved. 
    The conserved quantity ${\cal A}$ is analogous to the `angular momentum deficit', given by $\text{AMD} = \Gamma'_1+\Gamma'_2 = \Psi+\Phi$, which is a conserved quantity under purely secular evolution in the absence of MMRs \citep[e.g.,][]{Laskar1997,Laskar2000}.
    Roughly speaking, resonant planets conserve their `libration-averaged' AMD since ${\cal A}=\text{AMD}-kP$ and the resonant dynamics cause $P$ to oscillate about an equilibrium value $P\approx 0$.

    If, initially, $J^*={\cal A}$, then $\dot{J^*}=0$ for all time and the Hamiltonian can be reduced to
\begin{equation}
 {H}_{0}(J,\theta,J^*)=-\frac{A}{2k^2}(J-J^*)^2-\epsilon a_s J + \epsilon b_s J^*+H'_\text{res}(J,\theta)
    \label{eq:secular:integrable_ham}
\end{equation}
    so that the dynamics are completely integrable.
    In this case, the secular terms merely introduce a small shift to the center of the resonance through the `$\epsilon a_s J$' term. In general, when ${\cal A} > J^*$, the term
    \begin{equation}
            H_\text{pert}(J,\theta,J^*,\theta^*;{\cal A})= -\epsilon c_s\sqrt{J({\cal A} - J^*)}\cos(\theta+\theta^*)
            \label{eq:secular:pert_ham}
    \end{equation}
    couples the two degrees of freedom and $J^*$ varies with time. 

\begin{figure}
    \centering
    \includegraphics[width=\columnwidth]{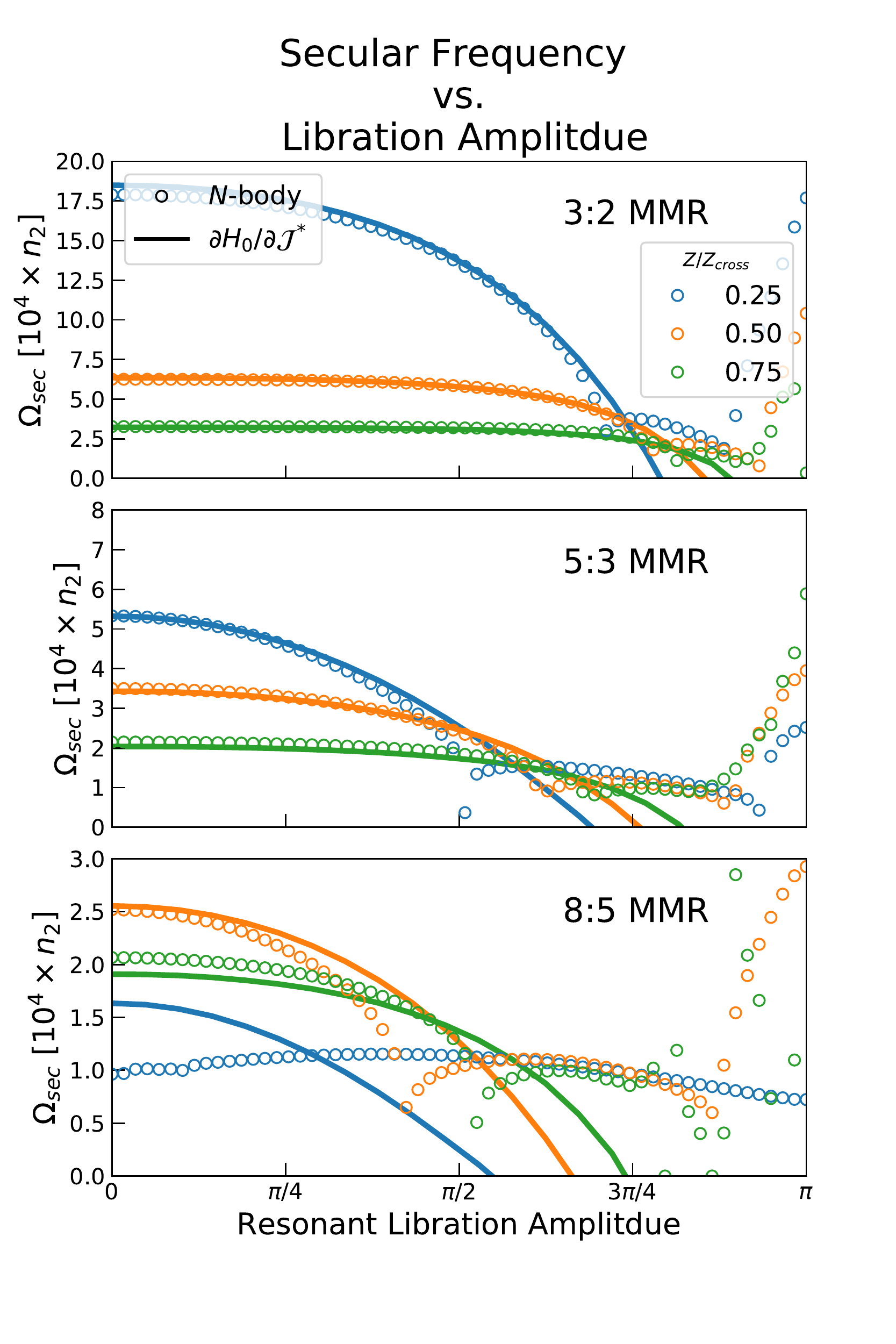}
    \caption{
    Absolute values of secular precession frequencies versus libration amplitudes of $k\theta$ for two $m_1=m_2=6\times10^{-6} M_*$ planets in various MMRs. 
    Conditions are chosen so that, initially, $W=0.125$, $w=0$, $a_1=\fracbrac{j-k}{j}^{2/3}a_2$, and $z=0$.
    The libration amplitude, $\delta \theta$, is set by initializing $\lambda_1=0$ and $\lambda_2=(\pi+\delta\theta)/j$.
    Different color curves correspond to different initial values of $Z$, measured in units of $Z_\text{cross}$, the value at which the orbits cross. 
    }
    \label{fig:secular:freq_vs_libration_amp}
\end{figure}

{
     To further analyze the secular dynamics of resonant planets, 
     I introduce the Arnold action-angle variables
    }
    \begin{align}
    {\cal J} = \frac{1}{2\pi}\oint Jd\theta~~&;~~\zeta =\frac{2\pi}{T_\text{res}}t \nonumber\\
    {\cal J}^* =J^* ~~&;~~
    \zeta^*=\theta^*+\rho({\zeta};{\cal J}^*,{\cal J})
    \label{eq:secular:action_angle_transformation}
\end{align}
{
    of the integrable Hamiltonian $H_0$, where $T_\text{res}$ is the resonant libration period, $t$ is time, and $\rho$ is a $2\pi$-periodic function of $\zeta$.
    I follow the semi-analytic procedure of \citet{Henrard1990} to affect this transformation \citep[see also][]{MorbidelliMoons1993,MoonsMorbidelli1995}.\footnote{Numerical routines for performing the transformation are available \online.}
    After the transformation to action-angle variables, the Hamiltonian $H_0({\cal J},{\cal J}^*)$ provides a mapping from the two action variables, ${\cal J}$ and ${\cal J}^*$, to two dynamical frequencies $\Omega_\text{res} \equiv \frac{\partial H_0}{\partial {\cal J}} = 2\pi/T_\text{res}$ and $\Omega_\text{sec} \equiv \frac{\partial H_0}{\partial {\cal J}^*}$.
    Figure \ref{fig:secular:freq_vs_libration_amp} compares values of $\Omega_\text{sec}=\frac{\partial H_0 }{ \partial{\cal J}^*}$, calculated with the semi-analytic procedure to frequencies measured from $N$-body simulations using the FMFT algorithm of \citet{FMFT1996}.
    The agreement between the semi-analytic and $N$-body results is quite good when $\Omega_\text{sec}$ is sufficiently far from zero. As I will show below, when $\Omega_\text{sec}\sim 0 $ the effect of $H_\text{pert}$ introduces a secondary resonance that is responsible for the the abrupt changes in the secular frequencies seen in the $N$-body results.
    } 

\begin{figure}
        \centering
        \includegraphics[width=\columnwidth]{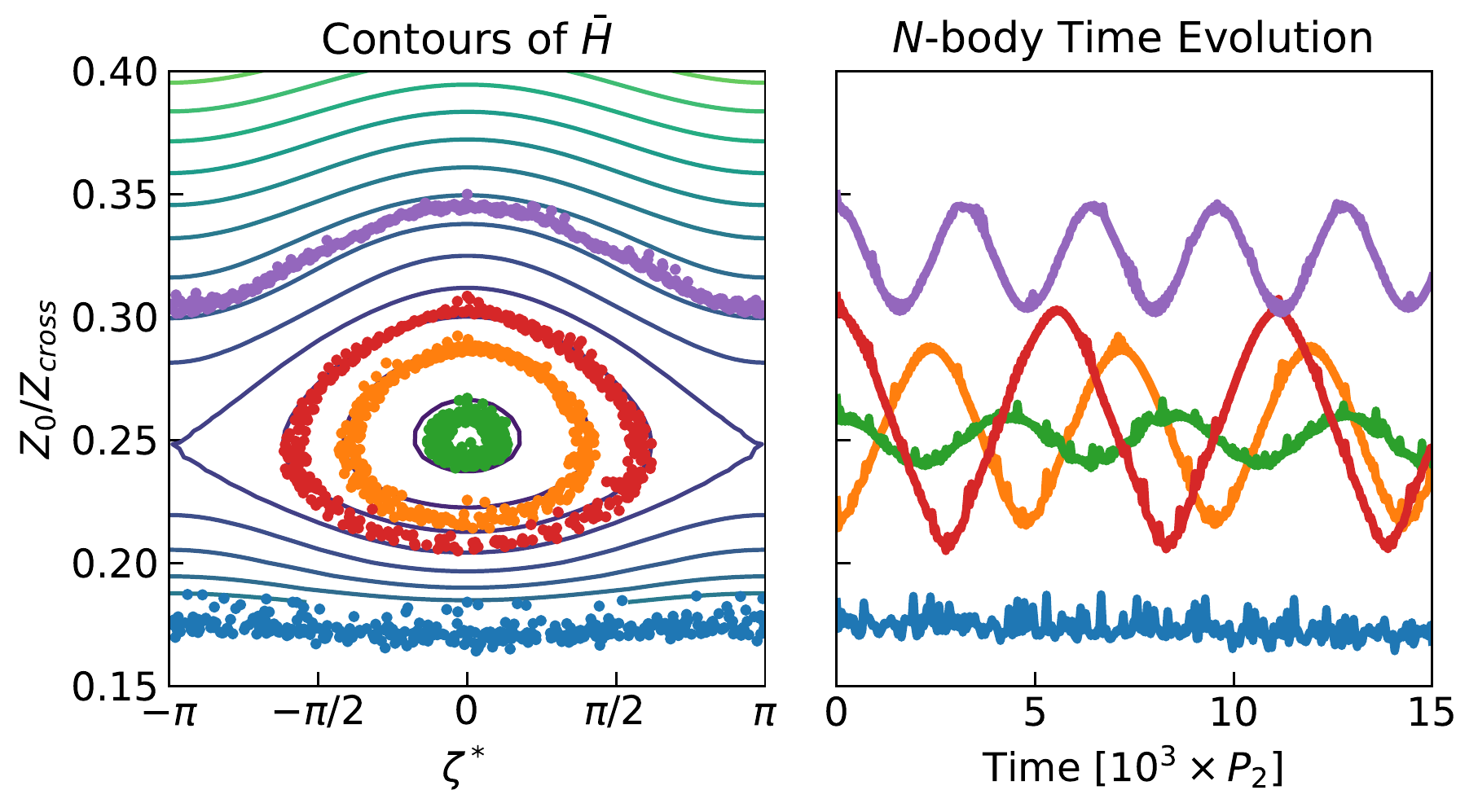}
        \caption{
        Left panel: contours of ${\bar H}$ (Equation \ref{eq:secular:sec_ham}) for a pair of planets in the 3:2 MMR with $m_1=m_2=3\times 10^{-5}M_*$. Contours are plotted in the $(\zeta^*,Z_0/Z_\text{cross})$ plane where I define $Z_0\equiv \fracbrac{f^2+g^2}{{\tilde f}^2+{\tilde g}^2}^{1/2}\sqrt{{\cal J}^*}$ by analogy with the relationship between $Z$ and $J$ given in Equation \eqref{eq:analytic_model:J_theta_def}.
        The adiabatic invariant ${\cal J}$ is chosen so that the $\Omega_{sec}=0$ resonance occurs at $Z_0/Z_\text{cross} = 0.25$ and ${\cal A}$ is chosen so that ${\cal A} = {\cal J}^*$ for $Z_0/Z_\text{cross}=0.5$.
        Results of $N$-body integrations are over-plotted as colored points.
        Right panel: the time evolution of $Z_0/Z_\text{cross}$ for the various $N$-body trajectories plotted in the left panel.
        Code reproducing this figure is provided \online.
        }
        \label{fig:secular:sos}
    \end{figure}

{
    After transformation \eqref{eq:secular:action_angle_transformation} the perturbing Hamiltonian becomes
    \begin{equation}
        H_\text{pert}=-\epsilon c_s\sqrt{({\cal A} -{\cal J^*})}
        \sum_{n=0}^{\infty}F_n({\cal J,J}^*)e^{i(\zeta^*+n\zeta)}
        + c.c.
    \end{equation}
    where the
    \begin{equation*}
       F_n({\cal J,J}^*) = \frac{1}{4\pi}\int_{0}^{2\pi}\sqrt{J(\zeta)}e^{i(\theta-\rho-n\zeta)}d \zeta
    \end{equation*}
    can be determined numerically \citep[see][]{Henrard1990}. 
   Thus, $H_\text{pert}$ introduces secondary resonances in regions of phase space where $\Omega_\text{sec} + n \Omega_\text{res} \approx 0$ for integer values of $n$. The secular frequency is typically significantly smaller than the libration frequency so that only the $n=0$ resonance condition can be satisfied outside of a narrow region near the separatrix where $\Omega_\text{res}\rightarrow 0$.  Accordingly, the secular dynamics are well-approximated by averaging $H_\text{pert}$ over $\zeta$ to obtain the Hamiltonian}
    \begin{equation}
    {\bar H} =  H_0({\cal J},{\cal J}^*) - \epsilon c_s \sqrt{{\cal A} - {\cal J}^*}|F_0({\cal J,J}^*)|\cos(\zeta^*)
    \label{eq:secular:hbar_secular}
    \end{equation}
    {
    describing an integrable, one degree-of-freedom system with ${\cal J}$ as a conserved adiabatic invariant. (\citet{BatyginMorbidelli2013analytic} also identify this adiabatic invariant in their treatment of first-order resonances, though they do not calculate it explicitly.) 
    Plotting contours of constant ${\bar H}(\zeta^*,{\cal J}^*;{\cal J},{\cal A})$ for a collection of trajectories with fixed ${\cal J}$ and ${\cal A}$, as in Figure \ref{fig:secular:sos}, can be used to obtain a global picture of the secular dynamics.  Note that far from the secular resonance ${\cal J}^*$ will oscillate about its mean value ${\cal J}^*_0$ with frequency $\approx \Omega_\text{sec}$ and amplitude $\approx {\epsilon c_s}\sqrt{{\cal J}^*_0({\cal A} - {\cal J}^*_0)}/\Omega_\text{sec}$, as can be shown from the equations of motion after noting  $|F_0|\approx \sqrt{{\cal J}^*}$.
    }
    
\begin{figure*}
    \centering
    \includegraphics[width=\textwidth]{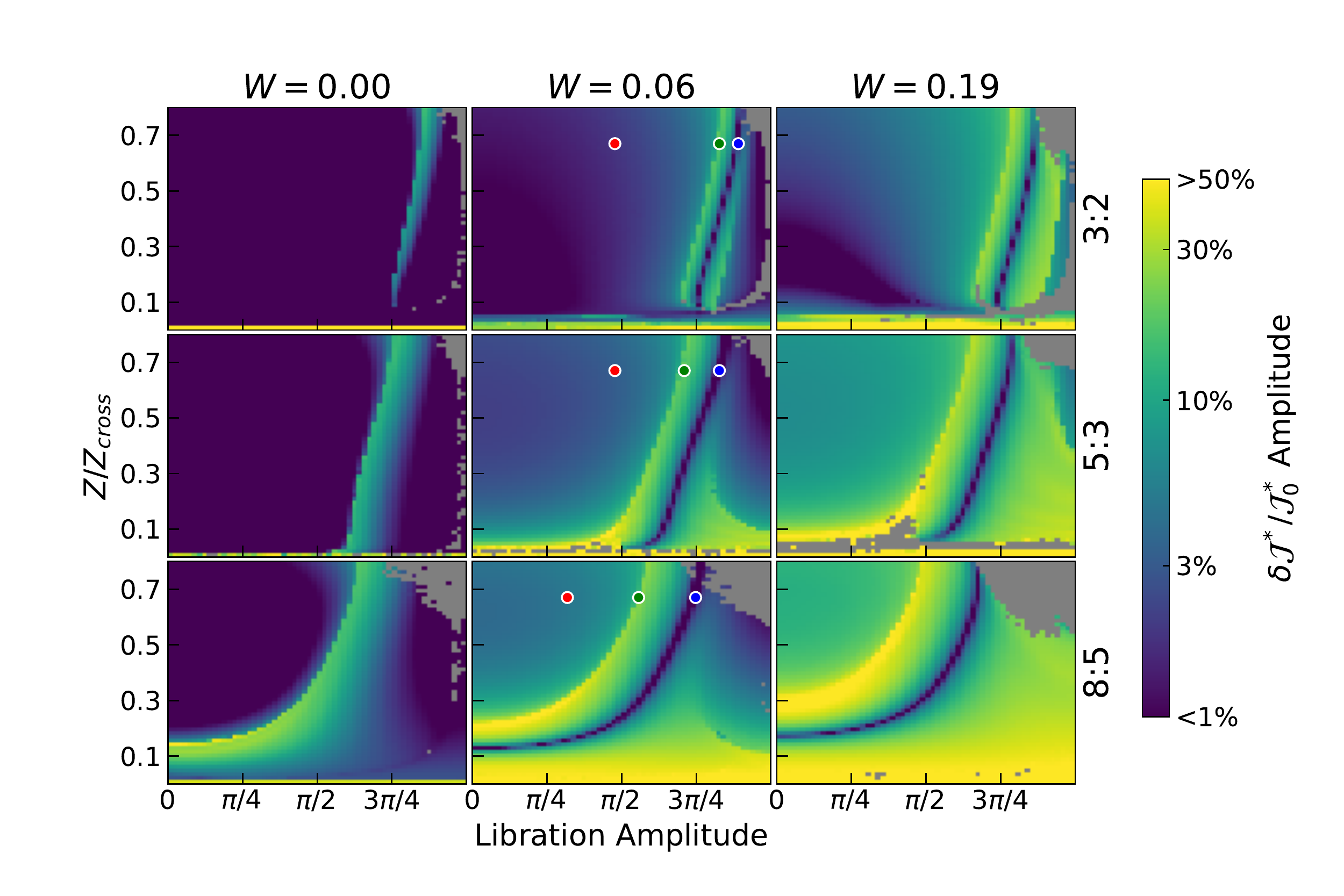}
    \caption{Maps of the amplitude of secular variations in ${\cal J}^*\propto Z^2$ determined from $N$-body simulations of two equal-mass $m_p=6\times10^{-6}M_*$ planets. Amplitudes are determined using the FMFT algorithm \citep{FMFT1996} from integrations spanning $1.8\times10^5$ orbits of the outer planet. Grid points are colored gray for initial conditions that resulted in significant chaos \citep[MEGNO$>10$;][]{Cincotta2003}. All other initial conditions are set in the same manner as Figure \ref{fig:secular:freq_vs_libration_amp}. Colored dots in the middle column indicate the initial conditions of the integrations plotted in Figure \ref{fig:secular:represenative_a_vs_Z}.}
    \label{fig:secular:amplitude_maps}
\end{figure*}

  Figure \ref{fig:secular:amplitude_maps} shows some representative maps illustrating the secular dynamics inside various resonances. 
  Each panel shows a color map of the magnitude of the fractional variation in ${\cal J}^*(t)$ measured from $N$-body simulations. 
  Secular resonances are evident in each map. 
  For the 3:2 and 5:3 MMRs with $W>0$, the secular resonance appear near libration amplitudes of $\sim 3\pi/4$ with ${\delta {\cal J}^*}/{{\cal J}^*_0}$ increasing from 0 at the center of the resonance to large values $\gtrsim 50\%$ towards the edges. 
  In the 8:5 MMRs with ${W}>0$, secular resonance occurs for all libration amplitudes at low $Z/Z_\text{cross}$. 

\begin{figure*}
    \centering
    \includegraphics[width=\textwidth]{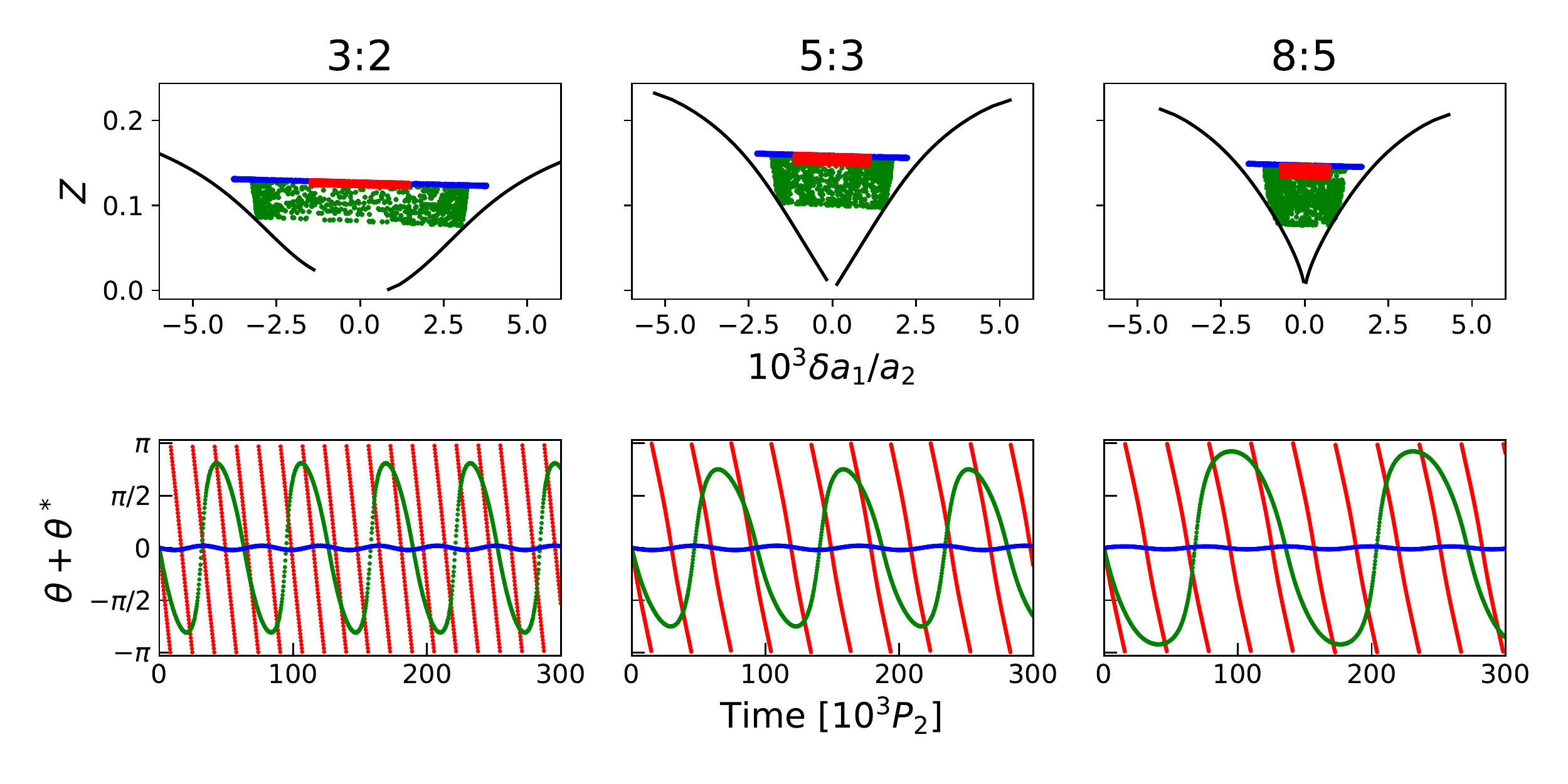}
    \caption{
    Secular evolution of resonant planets with different initial libration amplitudes. The top row shows planets' evolution in $\delta a_1/a_2$ versus $Z$ while the bottom row shows the time evolution of the critical angle, $\theta+\theta^*$, of the secular resonance.  Initial conditions are taken from Figure \ref{fig:secular:amplitude_maps} where the red, green, and blue points in the middle column show the initial conditions of the corresponding red, green, and blue plotted trajectories. Numerically-computed resonance separatrices, calculated as in Figure \ref{fig:numerical:res_grid}, are shown by black curves in the top row. 
    }
    \label{fig:secular:represenative_a_vs_Z}
\end{figure*}  

  Figure \ref{fig:secular:represenative_a_vs_Z} shows the secular evolution for some representative initial conditions indicated in the middle column of Figure \ref{fig:secular:amplitude_maps}.
  Red trajectories show the secular evolution for systems with moderate initial resonant libration amplitudes outside of the secular resonance. The critical angle of the secular resonance, $\theta+\theta^*$, circulates. These trajectories show small variations in the equilibrium $Z$ value ($\propto \sqrt{{\cal J}^*}$) about which resonant librations occur. Blue trajectories are selected deep in the secular resonance and show small oscillations about $\theta+\theta^*=0$ and again show minimal variations in their equilibrium $Z$ value. Green trajectories are selected to lie at the edge of the secular resonance and the secular critical angle, $\theta+\theta^*$, shows large-amplitude libration. These trajectories show large variations in their equilibrium $Z$ over the libration period of the secular critical angle while the amplitude of their oscillations in $\delta a_1/a_2$ remain roughly constant. As shown in top panels of Figure \ref{fig:secular:represenative_a_vs_Z}, these trajectories' decrease in $Z$ until, as a result of narrower resonance widths at lower $Z$ values, the oscillations in $\delta a_1/a_2$ nearly reach the resonance separatrix, at which point $Z$ increases again. 

  These numerical examples serve to illustrate that the secular evolution of resonant planet pairs exhibit a rich dynamics. A more thorough analysis of the secular evolution of resonant planets exploring the possibility of chaotic motion is beyond the scope of the present work.
  {
  However, before concluding I discuss some potential sources of secular chaos in MMRs identified by previous works. \citet{BatyginMorbidelli2013analytic} discuss the possibility for chaos caused by the resonant trajectory intermittently encountering the separatrix over the course of secular evolution. Such secularly-induced chaos is apparently largely absent from the numerical examples presented in this paper: chaotic regions of phase space, shown in gray in Figure \ref{fig:secular:amplitude_maps}, are relatively small and concentrated near the separatrix (i.e., libration amplitudes $\sim \pi$) at large eccentricities. Furthermore, it is unclear how much of this chaos is induced by secular modulations versus overlap with adjacent MMRs.\footnote{In \citet{BatyginMorbidelli2013analytic}'s example  exhibiting secularly-induced chaos, the planets are in the 2:1 MMR. Chaotic trajectories probably occupy a larger fraction of the phase space of this resonance compared to generic first-order MMRs because the 2:1 MMR disturbing function contains indirect terms that break the near-symmetry discussed in Section \ref{sec:why}.}
    \citet{MorbidelliMoons1993} and \citet{MoonsMorbidelli1995} present an extensive treatment of the secular dynamics of test-particles in MMRs with Jupiter and identify locations where resonant particles' apsidal precession frequencies are commensurate with the external forcing frequencies $\nu_5$ and $\nu_6$, the dominant harmonics in Jupiter's secular eccentricity evolution. 
    These secular resonances are analogous to the secular resonance identified here, though in the Jupiter-test particle case each MMR hosts two secular resonances corresponding to Jupiter's two secular modes.
    \citet{MorbidelliMoons1993} and \citet{MoonsMorbidelli1995}  show that these two secular resonances overlap and lead to wide chaotic regions inside Jupiter's MMRs.  This chaos destabilizes resonant orbits and is partially responsible for the formation of the Kirkwood gaps.
    Multiple secular resonances inside MMRs would likewise exist for resonant planet pairs in systems hosting more than two planets due to the presence of additional secular frequencies in the system. The overlap of such resonances could lead to chaos and potentially destabilize resonant planet pairs in multi-planet systems. 
    }

\section{Summary}
\label{sec:summary}
    I derived analytic (Equation \ref{eq:analytic_model:final_hamiltonian}) and semi-analytic (Equation \ref{eq:beyond_leading_order:Hfull}) integrable, one-degree-of-freedom Hamiltonian models for the dynamics of  MMRs of arbitrary order between co-planar planets. 
    Unlike many analytic models of MMRs, this model does not require that one of the planets be massless or on a circular orbit; the model only assumes co-planar planets on non-crossing orbits in or near a resonance interior to the 2:1 MMR.
    In Section \ref{sec:numerical}, I showed that this approximate model accurately describes the results of numerical integrations of resonant planet pairs.
    In Section \ref{sec:secular}, I examined the effects of longer-term secular evolution on the resonant dynamics.
    I showed that coupling between resonant planets' libration amplitude and secular frequency leads to a secular resonance in certain regions of phase-space that can cause significant modulations of the planets' eccentricities while elsewhere in phase space the secular evolution is characterized by small modulations of equilibrium eccentricities about which resonant oscillations occur. 

    The integrable model derived in this work provides a starting point for analyzing an array of dynamical phenomena such as resonant capture, resonances between mutually inclined planets, resonant chains, and secular evolution in systems of three or more planets containing resonant planet pairs.

\software{Matplotlib \citep{Matplotlib}, NumPy \citep{numpy}, REBOUND \citep{RL12}, WHFast \citep{RTwhfast2015}, SciPy \citep{scipy}, Theano \citep{theano}}
\acknowledgments
Acknowledgments.
I thank Yoram Lithwick for helpful conversations. I thank  Matt Payne for helpful comments on early drafts of this manuscript. I am grateful to Daniel Tamayo for helpful conversations and comments as well as his contributions to the {\it celmech} Python package utilized in this manuscript's numerical calculations.
SH gratefully acknowledges the CfA Fellowship.
The computations in this paper were run on the Odyssey cluster supported by the FAS Science Division Research Computing Group at Harvard University.
\bibliography{refs}

\appendix
\section{Fourier Representation of $R_\text{res}$}
\label{sec:appendix:fourier}
    In this Appendix I demonstrate some properties of the Fourier representation of the resonant disturbing function, $R_\text{res}$,  introduced in Equation \eqref{eq:analytic_model:Rres}. 
    I begin by writing the full disturbing function as a Fourier series in the planets' mean longitudes,
\begin{eqnarray}
        \frac{a_2}{|\vec{r}_2-\vec{r}_1|} &=& \sum_{p=-\infty}^{\infty}\sum_{q=-\infty}^{\infty} R_{p,q}(\alpha,z_1,z_1^*,z_2,z_2^*)e^{i(p\lambda_2+q\lambda_1)} + c.c.
        \label{eq:appendixFourier:fullDF}
\end{eqnarray}
    where $R_{p,q}(\alpha,z_1,z_1^*,z_2,z_2^*)$ are complex amplitudes. 
    {
    Rewriting the combinations of the $\lambda_i$ appearing in each Fourier term in terms of $\kappa$ and $Q$ introduced in Section \ref{sec:analytic_model:derivation} gives 
    \begin{equation*}
               p\lambda_2+q\lambda_1 =  (p (k-j)-j q)\kappa + \left(\frac{p}{k}+\frac{q}{k}\right)Q~.
    \end{equation*}
    We see that only Fourier terms for which $p (k-j)-j q = 0$ will give non-zero contributions when averaging the full disturbing function, Equation \eqref{eq:appendixFourier:fullDF}, over $\kappa$ to obtain $R_\text{res}$.
    Retaining only these terms from the double sum in Equation \eqref{eq:appendixFourier:fullDF} and subtracting $R_\text{sec}=R_{0,0}$}, the resonant disturbing function defined in Equation \ref{eq:analytic_model:Rres} is then
    \begin{eqnarray}
    R_\text{res}=\sum_{n=1}^{\infty} R_{nj,n(k-j)}(\alpha,z_1,z_1^*,z_2,z_2^*)e^{inQ} + c.c.~.
    \label{eq:appendixFourier:Rres}
    \end{eqnarray}    
        The D'Alembert property \citep[e.g.,][]{MD1999ssd,MorbidelliBook2002} implies that the amplitudes are of the form
\begin{eqnarray}
        R_{nj,n(k-j)}(\alpha,z_1,z_1^*,z_2,z_2^*) &=& \sum_{l=0}^{nk}\bigg\{C_{nj,nk,l}(\alpha)+\text{higher order terms}\bigg\}\times z_1^{*l}z_2^{*(nk-l)} 
\end{eqnarray}
    where the `higher order terms' are infinite series of monomials involving products of $|z_1|^{2}$, $|z_2|^{2}$, $(z_1^{*}z_2)$ and $(z_1z_2^{*})$ with coefficients that depend on $\alpha$.
    Because the complex eccentricities $z_1$ and $z_2$ are each a linear combinations of $Ze^{iz}$ and $We^{iw}$ (Eq. \ref{eq:analytic_model:Zdef}), the monomials $|z_1|^2,~z_1^{*}z_2,~z_1z_2^{*}$ and $|z_2|^2$ can be written as linear combinations  of the monomials $Z^2$ and $W^2$ and $ZWe^{\pm i(z-w)}$. Similarly,  $z_1^{*l}z_2^{*(k-l)}$ can be written as a linear combination of the monomials $(Ze^{-iz})^{k-l'}(We^{-iw})^{l'}$ with $0\le l' \le k$.
    Therefore 
\begin{eqnarray}
        R_{nj,n(j-k)}&=& \sum_{l'=0}^{nk}\bigg\{c_{nj,nk,l'}(\alpha)+\text{higher order terms}\bigg\}\times (Ze^{-iz})^{nk-l'}(We^{-iw})^{l'}  \label{eq:appendixFourier:RnjZW_I}
\end{eqnarray}
    where the new coefficients, $c_{nj,nk,l'}$, are linear combinations of the old $C_{nj,nk,l}$ and the `higher order terms' are infinite series of monomials involving products of $Z^2$ and $W^2$ and $ZWe^{\pm i(z-w)}$. 
    Let us rewrite Equation \eqref{eq:appendixFourier:RnjZW_I} as a generic Fourier series in $w$ and $z$,
\begin{equation}
R_{nj,n(j-k)} =
    \frac{1}{2}\sum_{l'=-\infty}^{+\infty}a_{n,l'}(Z,W)e^{-i(nk-l')z-il'w}~,
    \label{eq:appendixFourier:RnjZW_II}
\end{equation}
    where the sum now extends from $l'=\pm\infty$.
    Comparing terms in Equations \eqref{eq:appendixFourier:RnjZW_I} and \eqref{eq:appendixFourier:RnjZW_II}, we have, to leading order, that $a_{n,l'}\propto Z^{|nk-l'|}W^{|l'|}$.
    This is straightforward to see for $0\le l'\le k$. For $l'<0$ and $l'>k$ it is instructive to consider a concrete example: let's take $n=1,k=2$ and $l'=3$ corresponding to the Fourier term with complex phase $z-3w$ in Equation \eqref{eq:appendixFourier:RnjZW_II}. The lowest-order term with complex phase $z-3w$ appearing in Equation \eqref{eq:appendixFourier:RnjZW_I} (and thus the leading order term in a Taylor expansion of $a_{1,3}(Z,W)$) occurs in the $l'=2$ term of the sum as the product of $(We^{-iw})^{2} \times ZWe^{i(z-w)}$, the latter term occurring in the series of monomials represented as `higher order terms'. Thus, the amplitude $a_{1,3}\propto ZW^3 = Z^{|nk-l'|}W^{|l'|}$ to leading order as claimed.
    Indeed, the leading-order term all $a_{n,l'}$ for any $l'<0$ and $l'>k$ are formed by the product of $(Ze^{-iz})^{k-l'}(We^{-iw})^{l'}$, where $0\le l'\le k$, with and powers of $ZWe^{i(z \pm w)}$ occurring in the higher order terms of Equation \eqref{eq:appendixFourier:RnjZW_I}. 
    
\begin{figure}
        \centering
        \includegraphics[width=\textwidth]{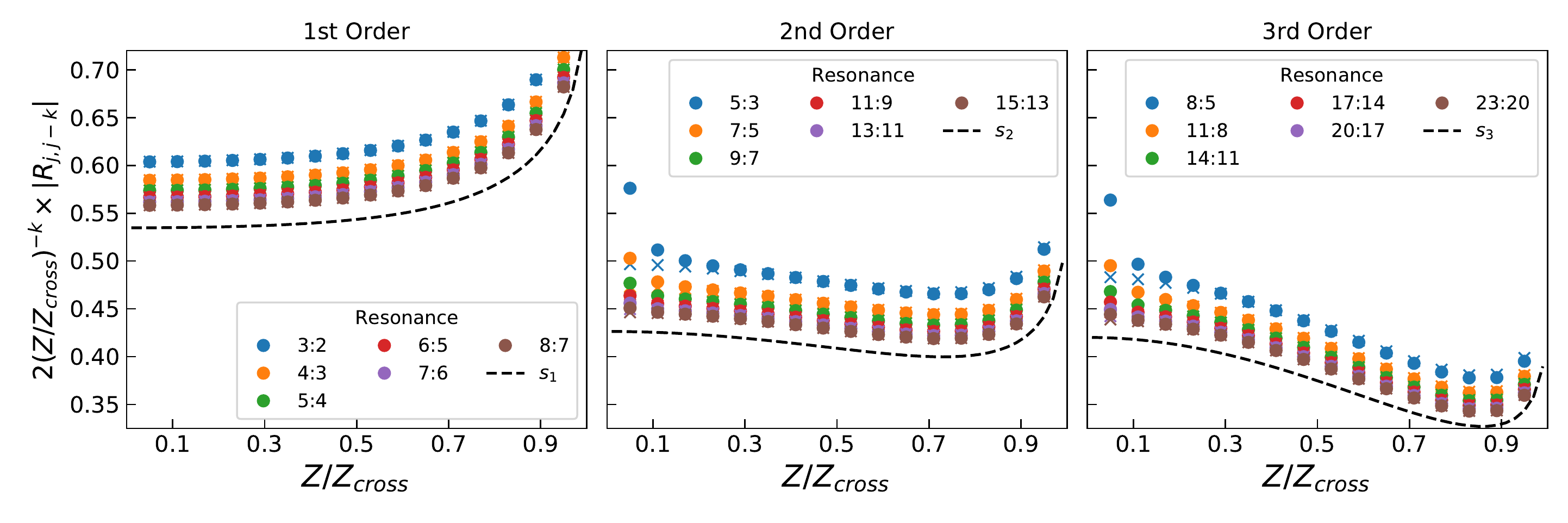}
        \caption{
        Scaling of Fourier coefficients $R_{nj,n(j-k)}$ with $Z/Z_\text{cross}$ for a series resonances. Each panel shows numerically-determined Fourier coefficients divided by $(Z/Z_\text{cross})^{k}/2$ where $k$ is the order of the resonance.
        Circles show coefficients evaluated with $W=0.3(\alpha^{-1}-1)$ and $w=z+\pi/2$ while `x's show coefficients evaluated with $W=0$. For most resonances and values of $Z/Z_\text{cross}$, the symbols lie on top of one another, confirming that $R_{nj,n(j-k)}$ is nearly independent of $W$. All resonances of the same order exhibit approximately the same functional dependence on $(Z/Z_\text{cross})$. Dashed lines show $s_k(Z/Z_\text{cross})$ defined in Equation \eqref{eq:appendixFourier:amplitude_approx}.
        }
        \label{fig:appendixFourier:amps_vs_zfrac}
    \end{figure}

    The results of Section \ref{sec:analytic_model:beyond_leading_order} can be restated as $R_{nj,n(j-k)} \approx  \frac{1}{2}a_{n,0}(Z,W)e^{-inkz}~.\label{eq:appendixFourier:RnjApprox}$
    The functional dependence of these Fourier amplitudes, $a_{n,0}$, on $Z$ and $W$ is specific to the particular $j$:$j-k$ MMR under consideration. 
     However, the  $a_{n,0}$ of distinct resonances of a given order $k$ are all well-approximated by a single function,
     \begin{eqnarray}
      a_{n,0}(Z,W) \approx s_{nk}(Z/Z_\text{cross})
      ={1\over\pi^2}
\int_0^{2\pi}K_0\left[\frac{2nk}{3}(1+(Z/Z_\text{cross})\cos M)\right]\cos\left[nk\left(M+\frac{4}{3} (Z/Z_\text{cross})\sin M\right)\right] {dM}~.\label{eq:appendixFourier:amplitude_approx}
     \end{eqnarray}
    where dependence on the particular resonance only enters only by determining the value of $Z_\text{cross}$.
    A similar result was derived in \citet{Hadden2018}, where it was shown that the cosine-amplitudes of disturbing function terms in the restricted three-body problem can be approximated by $s_{k}(e/e_\text{cross})$ where $e$ is the eccentricity of the test-particle and $e_\text{cross}$ is the orbit crossing eccentricity. 
    As this paper has demonstrated, the resonant dynamics of two eccentric planets is essentially identical to the dynamics of an eccentric test particle subject to a circular perturber after simply making the replacement $e/e_\text{cross}\rightarrow Z/Z_\text{cross}$ so the cosine amplitudes appearing in Equation \eqref{eq:analytic_model:Rres_full_approx} should match those in the restricted problem after making this same replacement. 
    Figure \ref{fig:appendixFourier:amps_vs_zfrac} compares numerically-computed $R_{nj,n(j-k)}$ coefficients with the approximation $|R_{nj,n(j-k)}|\approx \frac{1}{2} s_{nk}(Z/Z_\text{cross})$ and illustrates that \citet{Hadden2018}'s approximation for disturbing function coefficients readily generalizes to resonances in the unrestricted problem. The coefficients have been scaled by $(Z/Z_\text{cross})^{-k}$ to highlight deviations from the predicted leading-order behavior. 
    Two sets of $R_{nj,n(j-k)}$, one with $W=0$ and one with $W=0.3(\alpha^{-1}-1)$, are plotted for each resonance though the two values are indistinguishable in most instances, demonstrating that $R_{nj,n(j-k)}$ is nearly independent of $W$.
\section{The Secular Hamiltonian}
\label{sec:appendix:secular_derivation}
    Here I derive the Hamiltonian governing secular interactions, Equation \eqref{eq:secular:sec_ham}. The secular component of the disturbing function term,  to leading order in eccentricities, is given by
\begin{equation}
R_\text{sec} = f_2\left(e_1^2 + e_2^2\right) + f_{10}e_1e_2\cos(\varpi_2-\varpi_1)=\begin{pmatrix}z_1^*&z_2^*\end{pmatrix}\cdot\begin{pmatrix}f_2 & f_{10}/2\\f_{10}/2 & f_{2}\end{pmatrix}\cdot\begin{pmatrix}z_1\\z_2\end{pmatrix}\label{eq:appendix:Rsec}
\end{equation}
    where $f_{2}$ and $f_{10}$ are combinations of Laplace coefficients, defined in in Appendix B of \citet{MD1999ssd}.The secular Hamiltonian is given by $H_\text{sec}=-2 \epsilon R_\text{sec}$ after expressing $R_\text{sec}$ in terms of canonical variables. Equation \eqref{eq:analytic_model:PhiPsi_define} relates the complex eccentricities, $z_1$ and $z_2$, to the canonical variables introduced in Section \ref{sec:analytic_model} via the linear transformation
\begin{equation}
    \begin{pmatrix}
        z_1^*\\z_2^*
    \end{pmatrix} 
    =
    \begin{pmatrix}
        \sqrt{\frac{\mu_1+\mu_2}{\mu_1\sqrt{\alpha_0}}} & 0 \\
         0 & \sqrt{\frac{\mu_1+\mu_2}{\mu_2}}
    \end{pmatrix}
    \begin{pmatrix}
        \sqrt{\Gamma'_1}e^{i\gamma_1} \\ \sqrt{\Gamma'_2}e^{i\gamma_2}
    \end{pmatrix}
    =
    \frac{1}{\sqrt{\tilde{f}^2+\tilde{g}^2}}
    \begin{pmatrix}
        \sqrt{\frac{\mu_1+\mu_2}{\mu_1\sqrt{\alpha_0}}}\tilde{f} &     -\sqrt{\frac{\mu_1+\mu_2}{\mu_1\sqrt{\alpha_0}}}\tilde{g} \\
       \sqrt{\frac{\mu_1+\mu_2}{\mu_2}}\tilde{g} & \sqrt{\frac{\mu_1+\mu_2}{\mu_2}}\tilde{f}
    \end{pmatrix}
    \begin{pmatrix}
        \sqrt{\Phi }e^{i\phi} \\ \sqrt{\Psi}e^{i\psi}
    \end{pmatrix}~. \label{eq:appendix:complex_ecc_to_vars}
\end{equation}
    Inserting \eqref{eq:appendix:complex_ecc_to_vars} into Equation \eqref{eq:appendix:Rsec} and defining
        \begin{equation*}
            M\equiv                
                \frac{1}{\sqrt{\tilde{f}^2+\tilde{g}^2}}
            \begin{pmatrix}
                \sqrt{\frac{\mu_1+\mu_2}{\mu_1\sqrt{\alpha_0}}} & 0 \\
                 0 & \sqrt{\frac{\mu_1+\mu_2}{\mu_2}}
            \end{pmatrix}
                \cdot
                \begin{pmatrix}
                    \tilde{f} & -\tilde{g} \\
                    \tilde{g} & \tilde{f} 
                \end{pmatrix}~,
        \end{equation*}
    gives the disturbing function in terms of canonical variables as
        \begin{eqnarray}
        R_\text{sec}=
            \begin{pmatrix}
                    \sqrt{\Phi }e^{-i\phi} ,&
                    \sqrt{\Psi}e^{-i\psi}
            \end{pmatrix}        
            \cdot
            M^{T}\cdot
                \begin{pmatrix}
                    f_2 & f_{10}/2\\
                    f_{10}/2 & f_2
                \end{pmatrix}
                    \cdot
                    M             \cdot
            \begin{pmatrix}
                    \sqrt{\Phi }e^{i\phi} \\ 
                    \sqrt{\Psi}e^{i\psi}
            \end{pmatrix}        
        \end{eqnarray}
so that 
\begin{eqnarray}
    H'_\text{sec} &=& -\epsilon[ a_s \Phi + b_s \Psi + c_s\sqrt{\Phi \Psi}\cos(\phi - \phi)]\label{eq:appendix:hsec_PhiPsi}
\end{eqnarray}
where
\begin{eqnarray}
    a_s &=&\frac{2(\mu_1+\mu_2)}{{\tilde{f}^2+\tilde{g}^2}}\left[\paren{\frac{\tilde{f}^2}{\mu_1\sqrt{\alpha_0}}+\frac{\tilde{g}^2}{\mu_2}}f_2+\fracbrac{\tilde{f}\tilde{g}}{\sqrt{\mu_2\mu_1\sqrt{\alpha_0}}}f_{10}\right]\nonumber\\
    b_2 &=& \frac{2(\mu_1+\mu_2)}{{\tilde{f}^2+\tilde{g}^2}}\left[\frac{\tilde{f}^2}{\mu_1\sqrt{\alpha_0}}-\frac{\tilde{g}^2}{\mu_2}
    \right]f_{10}\nonumber\\
    c_s &=&\frac{2(\mu_1+\mu_2)}{{\tilde{f}^2+\tilde{g}^2}}\left[\paren{\frac{\tilde{f}^2}{\mu_1\sqrt{\alpha_0}}+\frac{\tilde{g}^2}{\mu_2}}f_2-\fracbrac{\tilde{f}\tilde{g}}{\sqrt{\mu_2\mu_1\sqrt{\alpha}}}f_{10}\right]~.
\end{eqnarray}
Finally, substituting $\Phi=J$, $\Psi = {\cal A}-J^*$, and $\phi-\psi=\theta+\theta^*$ in Equation \eqref{eq:appendix:hsec_PhiPsi} and dropping constant terms,
\begin{eqnarray}
H'_\text{sec}    =-\epsilon[ a_s J - b_s J^* + \sqrt{J ({\cal A}-J^*)}\cos(\theta + \theta^*)]~,  
\end{eqnarray}
which is the secular Hamiltonian introduced in Equation \eqref{eq:secular:sec_ham}.
\listofchanges
\end{document}